\documentclass[12pt]{article}
\usepackage{a4wide,epsfig}
\usepackage{scalefnt}
\usepackage{feynmf}

\usepackage[latin1]{inputenc}
\usepackage{amsmath}
\usepackage{amsfonts}
\usepackage[english]{babel}
\usepackage{slashed}

\newcommand{\nn}{\nonumber}
\newcommand{\be}{\begin{equation}}
\newcommand{\ee}{\end{equation}}
\newcommand{\bea}{\begin{eqnarray}}
\newcommand{\eea}{\end{eqnarray}}

\def\lQ{\Lambda_{\rm QCD}}

\newcommand{\mo}{\mathcal{O}}
\newcommand{\ns}{n\!\!\!/}

\newcommand{\Ds}{D\!\!\!/}
\newcommand{\as}{A\!\!\!/}
\newcommand{\vs}{v\!\!\!/}
\newcommand{\ps}{p\!\!\!/}

\newcommand{\la}{\left\langle}
\newcommand{\ra}{\right\rangle}
\newcommand{\lc}{\left[}
\newcommand{\rc}{\right]}
\newcommand{\lp}{\left(}
\newcommand{\rp}{\right)}
\newcommand{\bc}{\begin{center}}
\newcommand{\ec}{\end{center}}
\def\epm#1#2{\hbox{${\lower1pt\hbox{$\scriptstyle +~#1$}}
\atop {\raise1pt\hbox{$\scriptstyle -~#2$}}$}}
\newcommand{\lam}{\lambda}

\newcommand{\ml}{\mathcal{L}}

\newcommand{\xslash}[1]{{\rlap{$#1$}/}}
\def\dsl{\,\raise.15ex\hbox{/}\mkern-13.5mu D}
\def\siml{{\ \lower-1.2pt\vbox{\hbox{\rlap{$<$}\lower6pt\vbox{\hbox{$\sim$}}}}\ }}

\catcode`@=11
\newcount\@tempcntc
\def\@citex[#1]#2{\if@filesw\immediate\write\@auxout{\string\citation{#2}}\fi
  \@tempcnta\z@\@tempcntb\m@ne\def\@citea{}\@cite{\@for\@citeb:=#2\do
    {\@ifundefined
       {b@\@citeb}{\@citeo\@tempcntb\m@ne\@citea\def\@citea{,}{\bf ?}\@warning
       {Citation `\@citeb' on page \thepage \space undefined}}%
    {\setbox\z@\hbox{\global\@tempcntc0\csname b@\@citeb\endcsname\relax}%
     \ifnum\@tempcntc=\z@ \@citeo\@tempcntb\m@ne
       \@citea\def\@citea{,}\hbox{\csname b@\@citeb\endcsname}%
     \else
      \advance\@tempcntb\@ne
      \ifnum\@tempcntb=\@tempcntc
      \else\advance\@tempcntb\m@ne\@citeo
      \@tempcnta\@tempcntc\@tempcntb\@tempcntc\fi\fi}}\@citeo}{#1}}
\def\@citeo{\ifnum\@tempcnta>\@tempcntb\else\@citea\def\@citea{,}%
  \ifnum\@tempcnta=\@tempcntb\the\@tempcnta\else
   {\advance\@tempcnta\@ne\ifnum\@tempcnta=\@tempcntb \else \def\@citea{--}\fi
    \advance\@tempcnta\m@ne\the\@tempcnta\@citea\the\@tempcntb}\fi\fi}
\catcode`@=12

\begin{document}

\title{\vskip-3cm{\baselineskip14pt
\centerline{\normalsize UB-ECM-PF-06-13 \hfill}
\centerline{\normalsize  May 2006\hfill}
}
\vskip1.5cm
Heavy meson semileptonic differential decay rate 
\\
in two dimensions in the large $N_c$}
\author{Jorge Mond\'ejar,
Antonio Pineda\footnote{Permanent address after September 1st: Grup de F\'\i sica 
Te\`orica and IFAE, Universitat Aut\`onoma de Barcelona, E-08193 Bellaterra, 
Barcelona, Spain.}$\;$ and Joan Rojo\\[0.5cm]
{\small\it Dept. d'Estructura i Constituents de la Mat\`eria, U. Barcelona,}\\
{\small\it Diagonal 647, E-08028 Barcelona, Spain}}
\date{}

\maketitle

\thispagestyle{empty}

\begin{abstract}
We study QCD in 1+1 dimensions in the large $N_c$ limit 
using light-front Hamiltonian perturbation theory in the 
$1/N_c$ expansion. We use this formalism to 
exactly compute hadronic transition matrix elements for arbitrary currents 
at leading order in $1/N_c$.  
We compute the semileptonic differential decay rate of a heavy meson, 
$d\Gamma/dx$,  
and its moments, $M_N$, using the hadronic matrix 
elements obtained previously. We put some emphasis in trying to understand 
parity invariance. We also study with special care the kinematic region 
where the operator product expansion ($1/N \sim 1-x \sim 1$) or non-local 
effective field theories ($1/N \sim 1-x \sim \Lambda_{QCD}/m_Q$) can be applied. 
We then compare with the results obtained using an effective field theory approach 
based on perturbative factorization, with the focus to better understand quark-hadron duality.
At the end of the day, using effective field theories, we have been able to obtain expressions 
for the moments with relative accuracy of 
$O(\Lambda_{QCD}^2/m_Q^2)$ in the kinematic region where the operator product 
expansion can be applied, and 
with relative accuracy of $O(\Lambda_{QCD}/m_Q)$ in the kinematic region where non-local 
effective field theories can be applied. These expressions agree, within this precision, 
with those obtained from the hadronic result using the layer-function 
approximation plus Euler-McLaurin expansion.  
Very good numerical agreement for the moments is obtained between the exact result and 
the result using effective field theories.
\\[2mm]
\end{abstract}

\newpage


\tableofcontents

\vfill
\newpage

\section{Introduction}
\label{int}

Asymptotic freedom can be seen as the first example of 
factorization between high and low energies, since 
it dictates that Green functions 
at high Euclidean energies ($Q^2$) can be described by 
perturbation theory up to corrections suppressed by 
powers of $\lQ$ over $Q$. 
Therefore, the use of the operator product expansion (OPE) in processes where the
relevant momentum scale is large and Euclidean is safe. This is quite 
restrictive, since, in most of the cases, it can only be tested with experiment
through dispersion relations, which involve measurements up to
arbitrarily high energies. To avoid this problem what one usually does 
is to directly apply the 
same perturbative factorization techniques to observables living in the
Minkowski regime. In practice this means to perform the analytic continuation of 
approximate perturbative results obtained in the Euclidean region to the 
Minkowski region. Nevertheless, such calculations do not come from first 
principles. This affects the OPE and effective field
theories that are built using perturbative factorization techniques 
aiming to factorize high from low energies. This problem is
usually stated as duality violations. We will follow here the definition of
\cite{Shifman:2000jv} for duality violations. 

One can quantify the discrepancy between the exact result and the one 
using perturbative factorization in the large $N_c$ limit of QCD \cite{hooft1}. In 
this case one finds a clear discrepancy between both results 
in the physical cut of the Green functions, where one has infinitely 
narrow resonances on the one hand 
and an smooth function on the other. This can be further quantified in the 
't Hooft model \cite{hooft2}, which we will consider in what follows\footnote{One 
may believe that the large 
$N_c$ limit sets a kind of upper bound on the duality violations. In
the real case, the existence of finite decay widths is expected to smooth
the duality violations. However, at present, it is not possible to quantify 
this effect. In any case, this does not mean that one can use the results of the
two-dimensional model as upper bounds to the four-dimensional case with 
finite $N_c$.}. 
For this model, when one considers some inclusive 
quantities like the total heavy meson or 
tau decay rate,
the discrepancies between the hadronic and OPE-like result 
(using the OPE for the tau decay and HQET for the inclusive heavy 
meson decay \cite{Bigi:1998kc,lebed}) appear to be quite suppressed. 

On the other hand one may also study more exclusive quantities like 
the differential cross section of the electron-meson 
scattering going to electron+anything: $e M \rightarrow e X$ 
(deep inelastic scattering), the differential 
semi-leptonic inclusive decays of heavy mesons: $H_Q \rightarrow X l \nu$, 
or $e^+e^- \rightarrow$ light hadrons. Indeed one would 
expect that the magnitude of the duality violations for these 
quantities would be larger, since they are more exclusive observables. 
Some of them have been already studied in the literature \cite{callan,einhorn}, like 
deep inelastic scattering or $e^+e^- \rightarrow$ light hadrons. 
One finds that the violation of duality is maximal but that, if 
one makes some sort of smearing of the hadronic result, 
the partonic results are recovered at leading order. 
Nevertheless, for these observables, the partonic computation 
is performed at a diagrammatic level, which 
makes difficult to go beyond the leading order partonic result. 
In part this is so because one has to deal with jets (very energetic 
final states with relatively small invariant mass) in the final state. 
Moreover it is also 
difficult to quantify the error made by the smearing procedure, since 
the smeared function does not actually correspond to the differential 
cross section or decay anymore.

At this respect, there have been recent developments in order to apply  
effective field theories with perturbative factorization 
to jet physics in four dimensions. 
This is a rapidly evolving field \cite{Bauer:2000ew,bauerfact,Bauer:2001yt,Beneke1,neubert} 
and the effective field theory has been called soft-collinear effective theory 
(SCET). The use of effective field theories with perturbative 
factorization may allow for a comparison between partonic and hadronic 
results in a more systematic way, beyond the leading order partonic result,
and to set up a right 
framework on which to quantify the quark-hadron duality violations. Nevertheless, 
several questions still remain open in SCET, like what the modes of the 
theory are, or what the optimal formulation of this theory could be. The standard formulations 
of SCET involve the existence of a large number of modes, and one can never 
be sure that the set is complete. For instance, in Ref. \cite{neubert} it has been argued that there could 
be some extra modes called {\it messenger}. There is no consensus on this issue, 
though, and in Refs. \cite{Bauer:2003td,Manohar:2006nz} it is claimed that there is no need for such modes. 
Therefore, it is evident that 
the study of SCET is interesting on its own and the application of 
SCET in a controlled setup may help to better understand the structure of the 
effective theory. Obviously, QCD$_{1+1}$ 
provides this controlled setup. This will be one of the main subjects 
of this paper. On the side of the optimal formulation of SCET, we would like to 
incorporate the advantages of the light-front quantization frame \cite{Dirac:1949cp} 
and the associated Hamiltonian-like formulation. This provides 
non-trivial information, since it 
allows to relate the correlators that appear in the effective 
theory with the wave-function of the bound state. It also avoids 
to perform complicated diagrammatic computations and resummation 
via Dyson-Schwinger equations to obtain the 
matrix elements and vertices (as it was done in Refs. \cite{callan,einhorn}). Moreover, 
working in the light-cone gauge is convenient to make the theory
effectively abelian in 1+1 dimensions.

The specific observable we will consider in this paper to illustrate the discussion 
will be the differential semileptonic inclusive decays of heavy mesons: 
$H_Q \rightarrow X l \nu$. We will only consider the kinematical situation when 
the invariant mass square of the jet, $P_X^2$, is much larger than $\lQ^2 \sim \beta^2$ 
($\beta^2$ is the strong coupling, which has square mass dimensions in $D=1+1$).
In this 
situation we will see that one of the modes of SCET, the hard-collinear, is not 
a dynamical field and can be integrated out, at least in the 
light-front frame. The final effective theory becomes equal to HQET plus 
an imaginary vertex. This imaginary vertex is local in "time" 
(in the light-front quantization frame), can be computed order by order 
in perturbation theory, and is able to describe the differential decay 
rate (more precisely, the moments).

We structure the paper as follows. In sec. 2 we analyze QCD$_{1+1}$ in the 
light front and compute the transition matrix elements. In sec. 3 we 
compute the hadronic differential decay rate and moments. In sec. 4, we 
work out SCET in two dimensions and compute the differential decay rate 
and moments at tree level. In sec. 5, we develop an alternative effective 
theory without hard-collinear fields and compute the differential decay rate 
and moments at one loop. In the Appendix we set up the notation and conventions.

\section{QCD$_{1+1}$ in the light front}
\label{QCDLF}
In $D=1+1$, the QCD Lagrangian is given by
\be
\mathcal{L}_{1+1}=-\frac{1}{4}G^a_{\mu\nu}G^{a,{\mu\nu}}+
\sum_i \bar{\psi}_i\lp i\gamma^{\mu} D_{\mu}-m_i +i\epsilon \rp \psi_i
\,,
\ee
where $D_{\mu}=\partial_{\mu}+igA_{\mu}$ and $i$ labels the flavor. 

One can perform the quantization in a frame different from the equal-time frame.
In particular it is possible to choose the quantization frame at $x^+=constant$, which would play the
role of time in this case. The role of the energy is played by the 
conjugated variable $P^-$. The
other variables: $P^+$ (and $P_\perp$ in four dimensions) are kinematical. 
For instance, the $P^+_H$ component of an hadron behaves in "free"-particle way,
\be
P^+_{H}=\sum_i P^+_{i},
\ee
where the sum extends over all the partonic components of the bound state.
This allows to define the variable "x", which measures the fraction of
$P^+_{H}$ momentum carried by a given parton.

The notation for the components of the gluon field
is
\be
A^{+}\equiv n_+\cdot A, \quad A^{-}\equiv n_-\cdot A
\,.
\ee
The usual quantization gauge is $A^+(x)=0$. In this situation, the fields 
$\psi_-=\Lambda_-\psi$ (for the definition of $\Lambda_{+/-}$ see the appendix)
and $A^-$ are non-dynamical and can be integrated out from the 
theory (they are constraints)\footnote{One should not forget that there 
is another constraint, the Gauss law, that restricts the Hilbert space of 
physical states to those which are singlet under gauge transformations. See 
for instance \cite{Gaete:1993gk}, where one can also find a quantization in the path 
integral formulation.}. 
The resulting Lagrangian reads
($\psi_+=\Lambda_+\psi$)
\bea
\mathcal{L}=\sum_{i}\psi^{\dagger}_{i+}i\partial^-\psi_{i+}  + i\sum_i\frac{m_i^2-i\epsilon}{4}\int dy^-
\psi^{\dagger}_{i+}(x^-,x^+)\epsilon(x^--y^-)\psi_{i+}(y^-,x^+)
\nonumber \\ +\sum_{ij}\frac{g^2}{4}
\int dy^- \psi^{\dagger}_{i+}t^a\psi_{i+}(x^-,x^+)|x^--y^-|\psi^{\dagger}_{j+}t^a
\psi_{j+}(y^-,x^+) \   ,
\label{Lagqcd11}
\eea
where we have defined
\begin{equation}
\epsilon(x)=
\left\{
\begin{array}{ll}
-1\ , & x<0 \,,\\
0 \ , & x=0 \,,\\
1 \ , & x>0 \,.
\end{array}\right.
\\
\end{equation}
The representation of the quarks in terms of free 
fields in the light-cone quantization frame reads
\begin{equation}
\psi_+(x)=\int_0^{\infty}\frac{dp^+}{2(2\pi)}\left(
a(p) e^{-ipx}+
b^{\dagger}(p) e^{ipx}\right)
\,,
\end{equation}
and the anticommuting relations are
\begin{eqnarray}
&&\{a(p),a^{\dagger}(q)\}=\{b(p),b^{\dagger}(q)\}=2(2\pi) \delta(p^+-q^+) \,,
 \\ \nonumber
&&\{a(p),b^{\dagger}(q)\}=\{b(p),a^{\dagger}(q)\}=0
\,.
\end{eqnarray}

Once we have the Lagrangian we can construct the Hamiltonian (in the 
light-cone frame)
\begin{eqnarray}
\label{Pminus}
P^-=-i\sum_i\frac{m_i^2-i\epsilon}{4}\int dx^-dy^- \psi^{\dagger}_{i+}(x^-,x^+)
\epsilon(x^--y^-)\psi_{i+}(y^-,x^+)
\\
-\sum_{ij}\frac{g^2}{4}
\int dx^-dy^- \psi^{\dagger}_{i+}t^a\psi_{i+}(x^-,x^+)|x^--y^-|
\psi^{\dagger}_{j+}t^a
\psi_{j+}(y^-,x^+) 
\ . \nonumber
\end{eqnarray}

By solving the eigenstate equation (taking into account the 
constraints and where $n$ schematically labels the quantum numbers 
of the bound state)
\be
\label{eigenstate}
P^-|n\rangle=P_n^-|n\rangle
\,,
\ee
one obtains the basis of states on which the Hilbert space of physical 
states can be spanned. Here we will focus on the meson sector of the 
Hilbert space and we will generically label the state as $|ij;n\rangle$, 
where $i$ labels the flavor of the valence quark, $j$ labels the 
flavor of the valence antiquark and $n$ labels the excitation 
of the bound state.  
The solution to Eq. (\ref{eigenstate}) can be obtained from the large $N_c$ limit solutions 
 within a systematic 
expansion in $1/N_c$ using standard time-independent quantum perturbation 
theory. It has the following structure (the momentum of the bound state 
will not be displayed explicitly unless necessary)
\bea
&&
|ij;n\rangle=|ij;n\rangle^{(0)}
\\
\nn
&&
+
\sum_{m,n'}\sum_k|ik;n'\rangle^{(0)}|kj;m\rangle^{(0)}{}^{(0)}
\langle ik; n'|{}^{(0)}\langle kj;m|P^-|ij;n\rangle^{(0)}
\frac{1}{P_n^{(0)-}-P_m^{(0)-}-P_{n'}^{(0)-}}
\\
\nn
&&
+O\left(\frac{1}{N_c}\right)
\,,
\eea
where the second term in the expression is $1/\sqrt{N_c}$ suppressed. 
Here we have used the fact that, at order $1/\sqrt{N_c}$, $P^-$ only 
connects neighboring sectors ($n$-mesons $\rightarrow$ $n\pm 1$-mesons), 
becoming an almost diagonal infinite dimensional matrix. $|ij;n\rangle^{(0)}$
represents the eigenstate solution to Eq. (\ref{eigenstate}) in the large $N_c$
limit, and $P_n^{(0)}$ the associated eigenvalue (we do not explicitely 
display the flavor content of $P_n^{(0)}$ except in cases where it can 
produce confusion). In this limit the sectors 
with fixed number of quarks and antiquarks are conserved and consequently 
the number of mesons. Therefore, the bound state can be represented in the 
following way 
\begin{equation}
|ij;n\rangle^{(0)}=\frac{1}{\sqrt{N_c}}
\int_0^{P_n^+} \frac{dp^+}{\sqrt{2(2\pi)}}\phi^{ij}_{n}\left(\frac{p^+}{P_n^+}\right)
a_{i,\alpha}^{\dagger}(p)
b_{j,\alpha}^{\dagger}(P_{n}-p)|0\rangle
\,,
\end{equation}
where $\alpha$ is the color index, $\phi_{n}^{ij}$ is the solution to the 't Hooft equation,
which will be reviewed in the
next section, and the state is normalized as
\begin{equation}
{}^{(0)}\langle ij;m|i'j';n \rangle^{(0)}=2\pi 2P_n^{(0)+}\delta_{mn}\delta_{ii'}
\delta_{jj'}
\delta(P_m^{(0)+}-P_n^{(0)+})
\,.
\end{equation}

The fact that the number of particles is quasi-conserved makes possible 
to formulate the theory along similar lines of how is done in pNRQCD 
(for a review see \cite{Brambilla:2004jw}), where the wave function (the 't Hooft 
wave function in our case) is promoted to the status of being the 
field representing the bound state. 
We will not pursue this line of research further in this paper but we expect to come 
back to this issue in the future. 

\subsection{The 't Hooft equation}
\label{secthooft}
By applying the operator $P^-$  to its eigenstate $|n\rangle$ at leading 
order in $1/N_c$
one obtains the 't Hooft equation 
\be
\label{thoofteq}
M_n^2\phi^{ij}_n(x)={\hat P}^2\phi_n^{ij}(x)\equiv \lp \frac{m^2_{i,R}}{x}+ \frac{m^2_{j,R}}{1-x}\rp
\phi_n^{ij}(x)-\beta^2\int_0^1dy \phi^{ij}_n(y)\mathrm{P}\frac{1}{(y-x)^2}
\,,
\ee
where $M_n$ is the bound state mass, $\beta^2\equiv{g^2N_c\over 2\pi}$, $x=p^+/P_n^+$, with $p^+$ 
being the momentum of the 
quark $i$, and P stands for Cauchy's Principal Part\footnote{
One can use the following representation of this
distribution
\be
\label{pres}
P\frac{1}{(x-y)^2}=-\frac{1}{2}\int_{-\infty}^{\infty} dz |z|
e^{i(x-y)z}
\,.
\ee}. 
The renormalized mass is given by $m^2_{i,R}=m_i^2-\beta^2$. The
principal value prescription serves to regulate the integrand
singularity, which originates in the infrared divergence
of the gluon propagator. This equation has a discrete spectrum of
eigenvalues that increase approximately linearly for large $n$, and
the wave functions vanish at the boundaries with the asymptotic
behavior
\be
\label{endpoints}
\phi^{ij}_n(x)\to x^{\beta_i}, \quad x\to 0 \ ,
\ee
where
\be
m_{i,R}^2+\pi\beta_i\cot \pi\beta_i=0 \ ,
\ee
and similarly for $x\to 1$.
The 't Hooft wavefunction are chosen to be normalized to unity
\be
\int_0^1 dx \phi_n^{ij*}(x)\phi^{ij}_m(x)=\delta_{nm}
\,.
\ee
We will have to consider high excitations of mesons when  
considering the decay rate of the heavy meson. 
In the asymptotic limit $n\to \infty$ one can obtain
analytic expressions both for the masses
\be
M_n^2\simeq n\pi^2 \beta^2
\,,
\ee
and for the meson wave functions
\be
\label{wf_asym}
\phi^{ij}_n(x)\simeq\sqrt{2}\sin\lp n\pi x\rp
\,.
\ee
Actually, a more detailed study of the 't Hooft equation has been 
performed in the large $n$ limit using semiclassical (WKB approximation)
techniques in Ref. \cite{Brower:1978wm}. In this reference   
the layer function was defined (see also \cite{einhorn}):
\be
\phi_{i}(\xi) \equiv \lim_{n \rightarrow \infty} \phi^{ij}_n(\xi/M_n^2)
\,,
\ee
for finite $\xi$. This function is the solution of the equation
\be
\label{thoofteqlayer}
\phi_{i}(\xi)= \frac{m^2_{i,R}}{\xi}\phi_{i}(\xi)
-\beta^2\int_0^\infty d \xi' \phi_{i}(\xi')
\mathrm{P}\frac{1}{(\xi'-\xi)^2}
\,,
\ee
and the following equalities can be obtained 
\be
\int_0^{\infty} d\xi \frac{\phi_i(\xi)}{\xi}=\pi\frac{\beta}{m_i}
\,,
\qquad
\int_0^{\infty} d\xi \phi_i(\xi)=\beta\pi m_i
\,,
\ee
which we will need in the following sections.

\subsection{Transition matrix elements}
\label{tme}

We are now in the position to compute the transition matrix elements 
due to an arbitrary current:
\be
\label{matcurrent}
\langle c s ;m|{\bar \psi}_{c}\Gamma Q|Q s;n\rangle
\,,
\ee
where $\Gamma$ represents a generic Dirac matrix. 
We anticipate the notation that we will use for the heavy meson decay: 
$Q$ represents the field of the heavy quark as well as its flavor, 
$s$ the flavor of the spectator quark 
and $c$ the flavor of the hard-collinear quark. 
We will restrict ourselves to the kinematical situation relevant 
for the semileptonic heavy meson decay. This means that $P^+_m \leq P_n^+$ and 
$P^-_m \leq P_n^-$.  

\begin{figure}
\begin{center}
\includegraphics[width=0.49\columnwidth]{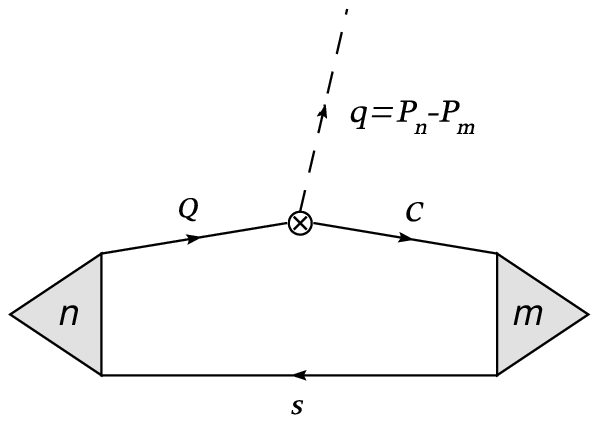}
\vskip1.5cm
\includegraphics[width=0.79\columnwidth]{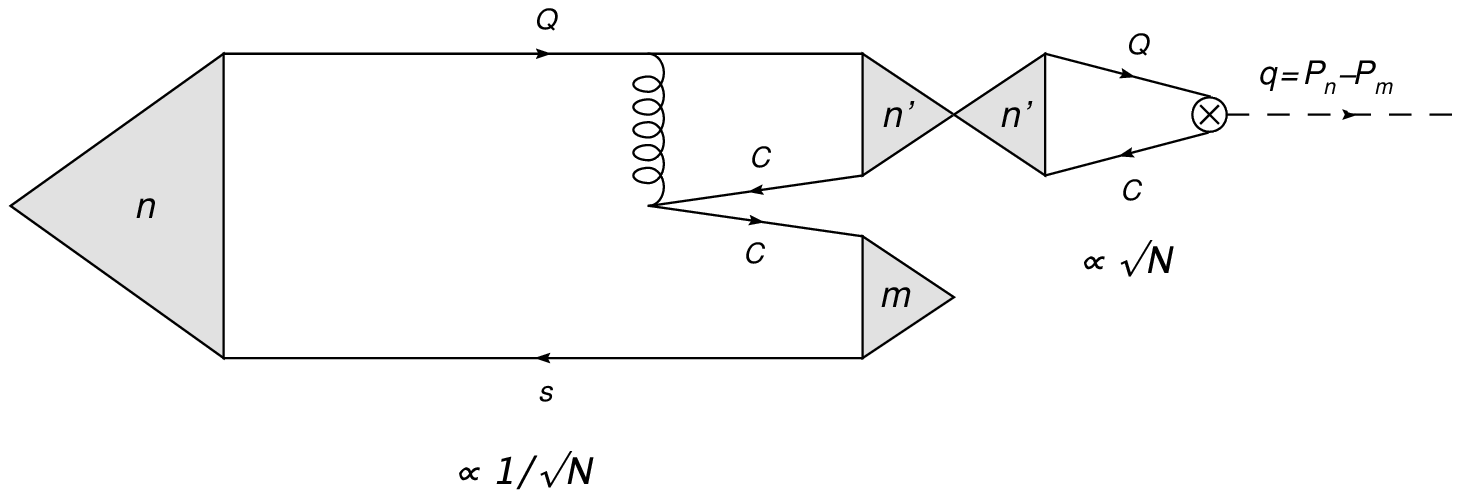}
\vskip1.5cm
\includegraphics[width=0.79\columnwidth]{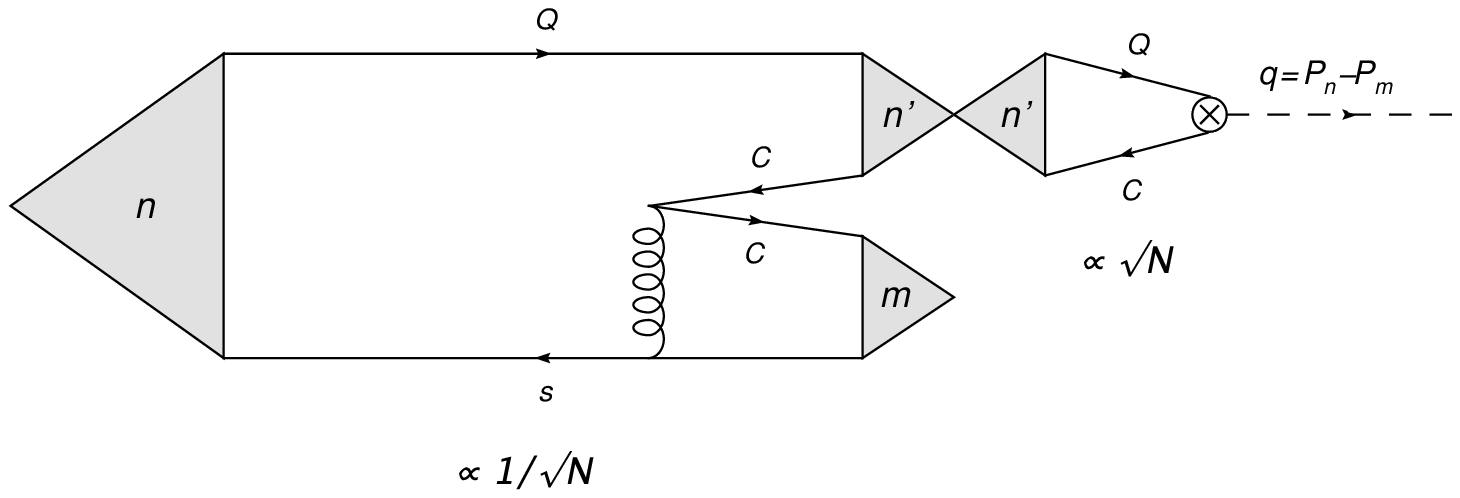}
\caption{\it Contributions to the hadronic matrix elements of the current. The first 
figure corresponds to the "diagonal" contribution to the matrix element, Eq. 
(\ref{currentdiag}). The second and third figures correspond to the 
"off-diagonal" terms, Eq. (\ref{offdiag}). The $\otimes$ represents the current, 
and the gluon exchange the effective four-fermion interaction in Eq. (\ref{Pminus}).}
\label{figmatelem}
\end{center}
\end{figure}

We only aim to obtain the matrix element (\ref{matcurrent}) at leading order in 
$1/N_c$. Nevertheless, this does not mean that we can just work with the leading 
order solution to the bound states. As we will see, we will also need the 
$1/\sqrt{N_c}$ corrections to the bound state. 
The contribution to the matrix element (\ref{matcurrent}) can be split into 
two contributions. We distinguish the contributions to the current 
according to whether they come from "diagonal" or "off-diagonal" terms, which 
we show in Fig. \ref{figmatelem}. The diagonal term directly connects the current to the 
leading $O(1/N_c^0)$ term of the bound state: 
\be
\label{currentdiag}
\langle c s ;m|{\bar \psi}_{c}\Gamma Q|Q s;n\rangle|_{diag.}
=
{}^{(0)}\langle c s ;m|{\bar \psi}_{c}\Gamma Q|Q s;n\rangle^{(0)}
\,.
\ee
This term is of $O(1/N_c^0)$ and it is produced from terms of the 
type ${\bar \psi}_{c}\Gamma Q \sim b_ca_Q^{\dagger}+\cdots$. In a way they 
change the flavor of the bound state from "heavy" to "hard-collinear". 
Nevertheless, there is 
another possibility: ${\bar \psi}_{c}\Gamma Q \sim b_cb_Q+\cdots$, which 
can be understood as the creation (annihilation) of a new bound state. This 
possibility does not have overlap with the leading order term in the 
$1/N_c$ expansion of the bound state but it does with the $1/\sqrt{N_c}$ one. 
Whereas the matrix element connecting the one meson sector with the two meson sector 
is $1/\sqrt{N_c}$ suppressed, the overlap of the two meson state with the 
current is $\sqrt{N_c}$ enhanced. This is why this contribution 
has to be considered as well at leading order in $1/N_c$.  We define 
\bea
\label{offdiag}
&&\langle c s;m|{\bar \psi}_{c}\Gamma Q|Q s;n\rangle|_{off-diag.}
=
\sum_{n'}\int \frac{dP_{n'}^+}{2(2\pi)P_{n'}^+}
\frac{1}{P_n^{(0)-}-P_m^{(0)-}-P_{n'}^{(0)-}}
\\
\nn
&&
\qquad
\times
\langle 0|{\bar \psi}_{c}\Gamma Q|Q c;n'\rangle^{(0)}
{}^{(0)}\langle Q c;n'|{}^{(0)}\langle c s;m| P^-|Q s ;n\rangle^{(0)}
\,.
\eea

A good thing of working this way is that, once 
${}^{(0)}\langle Q c;n'|{}^{(0)}\langle c s;m|P^-|Q s ;n\rangle^{(0)}$ 
has been computed, it can be used for any current. The total result 
for the matrix element at leading order in $1/N_c$ then reads
\be
\langle c s ;m|{\bar \psi}_{c}\Gamma Q|Q s;n\rangle
=
\langle c s ;m|{\bar \psi}_{c}\Gamma Q|Q s;n\rangle|_{diag.}
+
\langle c s;m|{\bar \psi}_{c}\Gamma Q|Q s;n\rangle|_{off-diag.}
\,.
\ee

We are now in the position to apply the above discussion to some 
specific currents. We display the results below ($x=q^+/P_n^+=
(P_n^+-P_m^+)/P_n^+$).
\bea
&&
\langle c s ;m|{\bar \psi}_{c}\gamma^+ Q|Q s;n\rangle
=
2\langle c s ;m|{\psi}^{\dagger}_{+,c} Q_+|Q s ;n\rangle
\\
&&
\quad
\nn
=
2P_n^+(1-x)
\left[
\int_0^1dz\phi^{Qs}_n(x+(1-x)z)\phi^{cs}_m(z)
\right.
\\
&&
\quad
\nn
\left.
-x^2
\beta^2
\int_0^1\int_0^1dudz
\frac{\phi^{cs}_m(z)G(u;q^2)}{(x(1-u)+(1-x)z)^2}(\phi^{Qs}_n(x+(1-x)z)-\phi^{Qs}_n(xu))
\right]
\,,
\eea
\be
G(u;q^2)\equiv \int_0^1 dv \sum_{n'=0}^{\infty}\frac{\phi^{Qc}_{n'}(u)\phi^{Qc}_{n'}(v)}
{q^2-M_{n'}^2}
\,.
\ee
\bea
\label{mel-}
&&
\langle c s ;m|{\bar \psi}_{c}\gamma^- Q|Q s ;n\rangle
=
2\langle c s ;m|\left(\frac{m_c}{i\partial^+}\psi_{c,+}\right)^{\dagger}
\lp\frac{m_Q}{i\partial^+}Q_+\rp|Q s ;n\rangle
\\
\nn
&&
=
\frac{2m_Qm_{c}}{P_n^+}\int_0^1dz\frac{\phi^{Qs}_n(x+(1-x)z)\phi^{cs}_m(z)}{(x+(1-x)z)z}
+2
\beta^2
\frac{1-x}{P_n^+}
\sum_{n'=0}^{\infty}\frac{(-1)^{n'}M_{n'}^2}{q^2-M_{n'}^2}
\\
\nn
&&
\qquad\qquad
\times
\int_0^1\int_0^1\int_0^1dydtdz
\frac{\phi^{Qc}_{n'}(y)\phi^{cs}_m(t)\phi^{Qc}_{n'}(z)}{(t(1-x)+(1-z)x)^2}
(\phi^{Qs}_n(x+(1-x)t)-\phi^{Qs}_n(xz))
\,.
\eea
\bea
\nn
&&
\langle c s ;m|{\bar \psi}_{c} Q|Q s;n\rangle
=
\int_0^1dz \phi^{Qs}_n(x+(1-x)z)\phi^{cs}_m(z)
\left(
\frac{m_Q(1-x)}{x+(1-x)z}+\frac{m_{c}}{z}
\right)
\\
&&-
\beta^2
\frac{x(1-x)}{m_Q-m_{c}}
\sum_{n'\,{\rm odd}}\frac{M_{n'}^2}{q^2-M_{n'}^2}
\\
\nn
&&
\qquad
\qquad
\times
\int_0^1\int_0^1\int_0^1dydtdz
\frac{\phi^{Qc}_{n'}(y)\phi^{cs}_m(t)\phi^{Qc}_{n'}(z)}{(t(1-x)+(1-z)x)^2}
(\phi^{Qs}_n(x+(1-x)t)-\phi^{Qs}_n(xz))
\,.
\eea
The "+" component of the vector current was already computed in Ref. 
\cite{einhorn}. The computation of the rest of the matrix elements had 
to wait to Ref. \cite{Burkardt:1991ea}, but we disagree with their results 
for the "-" component of the vector current and for the scalar current. 
More recently, expressions for the transition matrix elements of the 
current have also been worked out in Ref. \cite{Grinstein:1997xk}. 
We find that our expressions are more compact than those. In any case we 
have not been able to check the agreement with those. In Ref. \cite{Barbon:1994au} 
the matrix elements have also been considered using similar techniques 
to ours. Nevertheless, they consider different kinematics, which makes 
difficult the comparison with their results.

Finally, we would like to stress that current conservation imposes 
strong constraints on the form of the currents. The following equalities 
have to be fulfilled between the different matrix elements
\be
\label{currcons}
q_{\mu}
\langle c s ;m|{\bar \psi}_{c}\gamma^{\mu} Q|Q s;n\rangle
=(m_Q-m_c)
\langle c s ;m|
{\bar \psi}_{c} Q
|Q s;n\rangle
\,,
\ee
where $q^{\mu}=P_n^{\mu}-P_m^{\mu}$. 
For the case in which the hard-collinear and the heavy quark correspond 
to the same particle, and taking the limit $q^2, x \rightarrow 0$, we obtain 
the equality (first obtained in Ref. \cite{Burkardt:1989wy})
\be
\int_0^1\frac{(\phi^{ij}_{n})^2(x)}{x^2}
=
\frac{M_n^2}{m_{i,R}^2}
\int_0^1(\phi^{ij}_{n})^2(x)=\frac{M_n^2}{m_{i,R}^2}
\,.
\ee

\subsection{The static limit}

Since in this paper we will study the differential decay rate of a heavy meson, 
it is convenient to consider the specific case on which one of the quarks 
is very heavy (the static limit). If we redefine the heavy quark field,
\begin{equation}
Q_{+}=e^{-im_Qv\cdot x}Q_{+_v}
\,,
\end{equation}
where in the infinite mass limit one can use ($p=m_Qv + k$)
\begin{equation}
Q_{+_v}(x)=\int\frac{dk^+}{2(2\pi)} 
a_v(k) e^{-ikx}
\,,
\end{equation}
at leading order in $1/m_Q$ the Lagrangian reads
\begin{eqnarray}
\nn
\mathcal{L}_{static}&=&\sum_i\psi^{\dagger}_{i+}i\partial^-\psi_{i+} + i
\sum_i\frac{m_i^2-i\epsilon}{4}\int dy^- \psi^{\dagger}_{i+}(x^-,x^+)\epsilon(x^--y^-)
\psi_{i+}(y^-,x^+)  \\
\nn
&&+Q^{\dagger}_{+_v}(i\partial^-+i\partial^++i\epsilon)Q_{+_v}  
\\
\nn
&&
+\frac{g^2}{4}\sum_{ij}\int dy^-
\psi^{\dagger}_{i+}t^a\psi_{i+}(x^-,x^+)|x^--y^-|
\psi^{\dagger}_{j+}t^a\psi_{j+}(y^-,x^+) 
\\
&&+\frac{g^2}{2}\sum_{i}\int dy^-
\psi^{\dagger}_{i+}t^a\psi_{i+}(x^-,x^+)|x^--y^-|
Q^{\dagger}_{+_v}t^aQ_{+_v}(y^-,x^+)  
\,.
\end{eqnarray}
In the same way we can obtain the Hamiltonian in the static
limit (at leading order in the $1/m_Q$ expansion)
\begin{eqnarray}
P^-_{static}&=&-i\sum_i\frac{m_i^2}{4}\int dx^-dy^- \psi^{\dagger}_{i+}
(x^-,x^+)\epsilon(x^--y^-)\psi_{i+}(y^-,x^+) -Q^{\dagger}_{+_v}
(-i\partial^-)Q_{+_v}  
\nn
\\
\nn
&&
-\frac{g^2}{4}\sum_{ij}\int dy^-
\psi^{\dagger}_{i+}t^a\psi_{i+}(x^-,x^+)|x^--y^-|
\psi^{\dagger}_{j+}t^a\psi_{j+}(y^-,x^+) 
\\
&&-\frac{g^2}{2}\sum_{i}\int dy^-
\psi^{\dagger}_{i+}t^a\psi_{i+}(x^-,x^+)|x^--y^-|
Q^{\dagger}_{+_v}t^aQ_{+_v}(y^-,x^+)  
\,.
\end{eqnarray}
Once this Hamiltonian is applied to mesonic states, one obtains the
't Hooft equation in the static limit for which one can find a thorough 
study in Ref. \cite{Burkardt:2000ez}. In this limit, one works with the 
function $\Psi^{i}_n(t)=\frac{1}{\sqrt{m_Q}}
\phi^{Qi}_n\left(1-\frac{t}{m_Q}\right)$, where $t=(1-x)m_Q$, 
and considers its static limit, which is 
described by the following equation ($\epsilon_n=M_n-m_Q$, 
neglecting $1/m_Q$ 
corrections):
\be
\label{Hooftstatic}
\epsilon_n\Psi^{i}_n(t)=\frac{m_i^2-\beta^2}{2t}\Psi^i_n(t)+\frac{t}{2}\Psi^i_n(t)
-\frac{\beta^2}{2}\int_0^\infty ds \frac{\Psi^i_n(s)}{(t-s)^2}
\ .
\ee
The quantities that will be needed in the following 
 are expectation values
of the variable $t$ in the static limit,
defined in terms of the heavy meson wave function,
\be
\label{tavn}
\langle t^r \rangle\equiv \int_0^{\infty}dt \left( \Psi^{s}_n(t) \right)^2
t^r
\ .
\ee 
The numerical computation of these expectation values can be
cross-checked with the help of static limit
sum rules \cite{Burkardt:2000ez} such as
\be
\la t \ra=\epsilon_n \ ,
\ee
\be
\la t^2 \ra=\frac{4}{3}\la t \ra^2
-\frac{1}{3}\lp m_s^2-\beta^2\rp \ .
\ee
The above expressions will be useful in the
computation of the OPE expansions of the
differential decay rate moments, see sec. \ref{momhadr}.

\section{Semileptonic differential decay rate}
\label{SDDR}
\subsection{Kinematics}

We consider here the semileptonic heavy meson decay: $H_Q \rightarrow X_c l_a \bar l_b$, 
where $H_Q$ represents a bound state made of a heavy quark $Q$ and a light (spectator) 
quark $s$ (by default we will perform the numerical analysis for the ground 
state but the formulas hold for any state). $X_c$ represents any hadronic final state with $c$ (hard-collinear) 
flavour content and $l_{a,b}$ represent massless leptons. We will consider the situation on which 
the spectator, $\psi_{s}$, and hard-collinear, $\psi_c$, quarks have different flavour in order to avoid annihilation and Pauli 
interference terms. 
This decay has already been studied in the past. We will follow here the work 
of Bigi et al. \cite{Bigi:1998kc}. The authors considered the flavour 
changing weak interaction 
\be
\label{lagweakhad}
{\cal L}_{\rm weak}^V=-\frac{G}{\sqrt{2}}\bar \psi_c \gamma_{\mu}Q \bar
l_a \gamma^{\mu} l_b
\,.
\ee
The total decay width can be written 
as
\be
\Gamma_{H_Q}=\frac{G^2}{M_{H_Q}}\int \frac{d^2q}{(2\pi)^2} \theta(q^+)\theta(q^-)
Im \Pi_{\mu\nu}(q)
Im T^{\mu\nu}(q)
\,,
\ee
where $\Pi_{\mu\nu}(x)$ and $ T_{\mu\nu}(x)$ are defined as
\begin{equation}
  \label{Pi1}
  \Pi_{\mu\nu}(x)= i\,\langle 0|T \left \{ \bar{l}_a(x) \gamma_\mu l_b(x)
\, \bar{l}_b(0) \gamma_\nu l_a(0)\right\}|0\rangle\;,
\end{equation}
\begin{equation}
  \label{T1}
T^{\mu\nu}(x)= i\,\langle H_Q |T \left \{ \,
\bar{Q}(x) \gamma^\mu \psi_c(x) \,\bar{\psi}_c(0) \gamma^\nu Q(0)\right\}|H_Q\rangle\;,
\end{equation}
and their Fourier transform as 
\begin{equation}
\Pi_{\mu\nu} (q)=\int {\rm d}^2 x \, e^{iqx} \Pi_{\mu\nu}(x)
,\qquad
T^{\mu\nu}(q)=\int d^2x e^{-iqx}T^{\mu\nu}(x)
\,.
\label{piq}
\end{equation}
The imaginary part of $\Pi_{\mu\nu} (q)$ reads
\begin{eqnarray}
\label{ImPi}
Im \Pi^{\mu\nu} (q)=q^\mu q^\nu \delta(q^2)=
\,\left\{
\begin{array}{ll}
& q^+\delta(q^-) \quad (+,+) \;{\rm component},
\\
& q^-\delta(q^+) \quad (-,-) \;{\rm component},
\\
& 0 \quad\quad \quad \quad{\rm otherwise}\,.
\end{array} \right.
\end{eqnarray}
We notice that the result Eq. (\ref{ImPi}) is the same in the equal-time 
or in the light-front formalism. This is so in the massless case. Once 
masses are included the situation becomes more complicated. 
One may wonder how it is possible to obtain the term proportional 
to $\delta(q^+)$ and {\it finite $q^-=P_a^-+P_b^-=m_a^2/P_a^++
m_b^2/P_b^+$} in the leptonic correlator 
for massless leptons. This is somewhat amusing if one works in the light-front 
quantization frame. Naively one would expect that $q^-$ is always zero 
if the masses of the leptons are zero.  Nevertheless, one may obtain 
$Im\Pi^{--}=q^-\delta(q^+)$ (which is 
necessary to restore parity invariance) by working with 
finite masses for the leptons and taking the massless limit at the 
very end. Being more precise, this contribution appears from very 
high $P_{a,b}^+$, scaling like $P_{a,b}^+ \sim \tilde P_{a,b}^+/m_{a,b}^2$
with $\tilde P_{a,b}^+$ finite.

We can see that two terms are generated for the differential decay rate
(where $M_{H_Q}$ is the mass of the $H_Q$ meson):
\be
\frac{d\Gamma^{(+)}}{dx} \equiv \frac{G^2M_{H_Q}}{2(4\pi)^2}xImT^{--}(q^-=0,q^+)
\ee
with $x=q^+/M_{H_Q} \geq 0$, and 
\be
\frac{d\Gamma^{(-)}}{dx} \equiv \frac{G^2M_{H_Q}}{2(4\pi)^2}xImT^{++}(q^-,q^+=0)
\ee
with $x=q^-/M_{H_Q} \geq 0$. The total decay width then reads
\be
\Gamma_{H_Q}=\int_0^1 dx \left(\frac{d\Gamma^{(+)}}{dx}
+
\frac{d\Gamma^{(-)}}{dx} \right)
\,.
\ee
 
The procedure of Ref. \cite{Bigi:1998kc}
was to assume that
\be
\label{GmGm}
\frac{d\Gamma^{(+)}}{dx}=\frac{d\Gamma^{(-)}}{dx}
\equiv
\frac{1}{2}\frac{d\Gamma}{dx}
\ee
are equal by parity symmetry 
and only to compute $\frac{d\Gamma^{(-)}}{dx}$. That both terms are equal is indeed 
highly non-trivial due to the fact that the gauge fixing $A^+=0$ and working 
in the light-front quantization frame breaks the explicit invariance under 
parity. We will here explicitly compute $\frac{d\Gamma^{(+)}}{dx}$ and compare with 
$\frac{d\Gamma^{(-)}}{dx}$. We then have to compute the differential decay rate (in Ref. 
\cite{Bigi:1998kc}
only the total decay rate was considered). 
As we will see, its computation is highly non trivial and 
requires to take the massless limit for the hard-collinear 
quark in a careful way. 

\begin{figure}[h]
\begin{center}
\includegraphics[width=0.49\columnwidth]{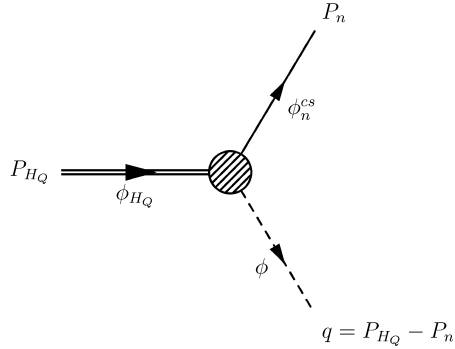}
\caption{\it Decay of the heavy meson $H_Q$ to the meson 
$|cs;n\rangle$ and the fictitious $\phi$ particle.}
\label{bgamma}
\end{center}
\end{figure}

Eq. (\ref{ImPi}) shows that, at the practical level, 
the interaction could be simulated by a massless particle. 
Therefore, there are some kinematical 
similarities with a kind of $b \rightarrow X_c \gamma$ decay. 
The effective interaction would read \cite{Bigi:1998kc}
\be
{\cal L}^V_{\rm weak}=-{G \over \sqrt{2\pi}}{\bar \psi}_c\gamma_{\mu}Q\epsilon^{\mu\nu}(\partial_{\nu}\phi)
+(h.c.)
\,,
\ee
where $\phi$ represents a fictitious pseudoscalar particle. 
The process can be seen in Fig. \ref{bgamma}. Since the pseudoscalar particle is
real,
we can parametrize its two-momentum $q$ as (for definiteness we set the kinematics 
relevant for the computation of $\frac{d\Gamma^{(+)}}{dx}$, for $\frac{d\Gamma^{(-)}}{dx}$ 
things work analogously)
\be
q^0=\frac{xM_{H_Q}}{2}, \qquad q^1=\frac{xM_{H_Q}}{2}, \qquad
q^-=0,
\ee
where, by momentum conservation, the momentum of the final hadronic state 
$P_X=P_n$ reads
\be
P_{X}^0=\frac{M_{H_Q}}{2}(2-x), \quad
P_{X}^1=-q^1, \quad P_X^2=M_{H_Q}^2(1-x)
\,.
\ee
In light-cone coordinates
\be
P_{X}^-=P_{X}^0-P_{X}^1=M_{H_Q}, \quad P_{X}^+=P_{X}^0+P_{X}^1=M_{H_Q}(1-x)
\,,
\ee
so we see that in the endpoint region $x\to 1$, the factor
\be
\sqrt{1-x}=\sqrt{P_X^2 \over M_{H_Q}^2}\equiv \bar \lambda
\ee
is small. It will play the role of one of the SCET expansion parameter, and leads
to appropriate scalings for the momentum of the final meson state:
\be
P_X^-\sim 1, \quad P_X^+\sim {\bar \lambda}^2 \ ,
\ee
which behaves as a collinear jet.

The heavy quark mass will be considered to be a large parameter 
(equivalent to $Q$ in jet physics or deep inelastic scattering). 
We can distinguish at least three kinematical regimes 
(we use the names SCETI and SCETII for an easier comparison with 
the notation used in effective field theories, see next sections and 
Ref. \cite{Bauer:2002aj}):
\begin{itemize}
\item[a)]
OPE; $P_{X}^2=M_n^2=M_{H_Q}^2(1-x) \sim m_Q^2 \gg m_Q\lQ$; $n \sim m_Q^2/g^2$,
\item[b)]
SCETI; $P_{X}^2=M_n^2=M_{H_Q}^2(1-x) \sim  m_Q\lQ$; $n \sim m_Q/g$,
\item[c)]
SCETII; $P_{X}^2=M_n^2=M_{H_Q}^2(1-x) \sim \lQ^2$; $n \sim 1$.
\end{itemize}

One usually refers to situation c) as the most exclusive and a) as the less
exclusive one. Here we would like to stress that in the large $N_c$ limit
(irrespectively of the number of spatial-time dimensions) all the
three situations are equally exclusive, since they correspond to only one
physical hadronic final state. $n$ represents the principal quantum number of
the hadronic excitation (we are having in mind a linear Regge behavior).
Indeed, the jet multiplicity of the hadronic final state 
is not well represented in the large $N_c$.

We have two independent expansion parameters: 
$\lambda=\sqrt{\Lambda \over M_{H_Q}}$ and $\bar \lambda=\sqrt{1-x}$. 
We note that $\lambda << 1$ is always fulfilled. With respect $\bar 
\lambda$, we will restrict ourselves to the situation where we are either 
in the OPE or SCETI situation. 

Let us note at this stage that we are actually using the opposite 
kinematic condition to the one used in Ref. \cite{Bigi:1998kc}. There 
is a reason for that. In our kinematics, the 
"energy" $P^-_X$ of the hadronic jet is much larger than $\lQ$. Therefore,
in the "time"-axis, $x^+$, the interaction takes place at very short 
times and can be considered local. This is what it will allow us to 
write the interaction as a local term (in "time", i.e. $x^+$) when we 
try to represent the process by means of effective field theories later on. 
See Fig. \ref{Bmatching}.
This will also allow us to write the matrix elements in terms of the wave-function 
of the bound state. We will elaborate on this in secs. 4 and 5.

\subsection{Differential decay rate: hadronic computation}

We can use the spectral decomposition to relate Im$T$ with the transition 
matrix elements of the currents we computed in sec. \ref{tme}. We obtain
(we have already restricted to $q^+$, $q^- \geq 0$)
\bea
&&
Im T^{\mu\nu}(q)= (2 \pi)^2\sum_n 
\int \frac{dP_n^+}{(2\pi)2P_n^+}
\delta(-q^++P_{H_Q}^+-P_n^+)\delta(-q^-+P_{H_Q}^--P_n^-)
\nn
\\
&&
\times
\langle H_Q|\bar{Q}(0) \gamma^\nu \psi_c(0)|cs;n,P_n^+\rangle
\langle cs; n,P_n^+|\bar{\psi}_c(0) \gamma^\mu Q(0)|H_Q \rangle
\,.
\eea

The expression for the differential decay rate then reads 
(we work in the rest frame of the heavy meson with $P_{H_Q}^+=P_{H_Q}^-=M_{H_Q}=P_n^-$ 
and $x=1-P_n^+/P_{H_Q}^+=1-M_n^2/M_{H_Q}^2$)
\be
\label{dGdx+}
\frac{d\Gamma^{(+)}}{dx}=\frac{G^2M_{H_Q}}{32\pi }\sum_{M_{n}\le M_{H_Q}}
\frac{x}{P_{H_Q}^+(1-x)}\Big| \langle n;P_n^+|\bar{\psi}_c(0) \gamma^- Q(0)|H_Q \rangle\Big|^2
\delta\lp P_{H_Q}^--P_n^-\rp
\,,
\ee
where the matrix element can be read from Eq. (\ref{mel-}) taking the limit $q^2 \rightarrow 0$.
We notice that the differential decay rate consists of a sum over deltas at the position of the 
resonances and, therefore, cannot be obtained from perturbative-like computations. 

\begin{figure}[h!!]
\makebox[1.0cm]{\phantom b}
\put(-30,10){\epsfxsize=8truecm \epsfbox{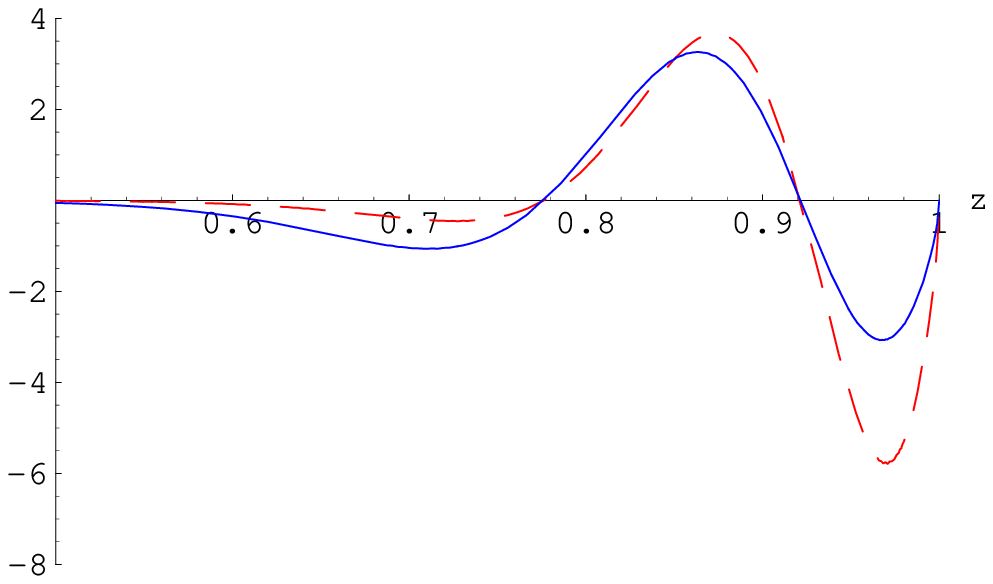}}
\put(210,10){\epsfxsize=8truecm \epsfbox{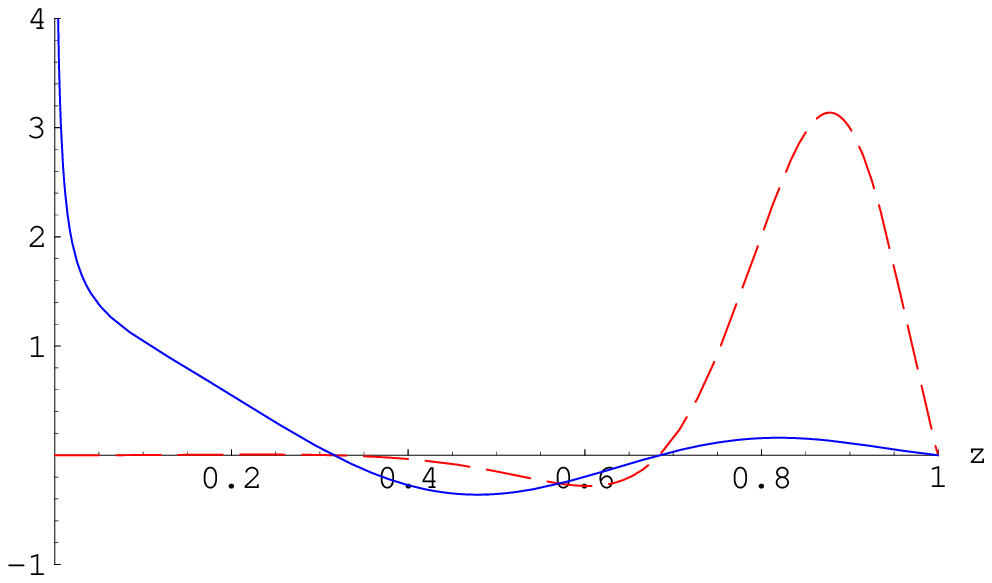}}
\caption{\it Plot of the integrands of Eq. (\ref{equality}). 
We have taken the values $n=3$ and $m_c=10\beta$, $m_{s}=0.56\beta$ and 
$m_Q=15\beta$ in the first figure and $m_c=1\beta$, $m_{s}=1\beta$ and 
$m_Q=10\beta$ in the second figure. The dashed red line corresponds to the 
integrand of the left-hand side of the equality. The solid blue line 
corresponds to the integrand of the first term (the second term is subleading in 
$1/m_Q$ and it is not considered in this plot) of the 
right-hand side of the equality. In the first figure the solid 
blue line 
diverges (although in an integrable manner) 
for $z \rightarrow 0$ but it cannot be seen with the resolution of the 
plot.}
\label{supportplots} \vspace{1mm}
\end{figure}

\begin{figure}[h!!]
\begin{center}
\makebox[1.0cm]{\phantom b}
\epsfxsize=10truecm \epsfbox{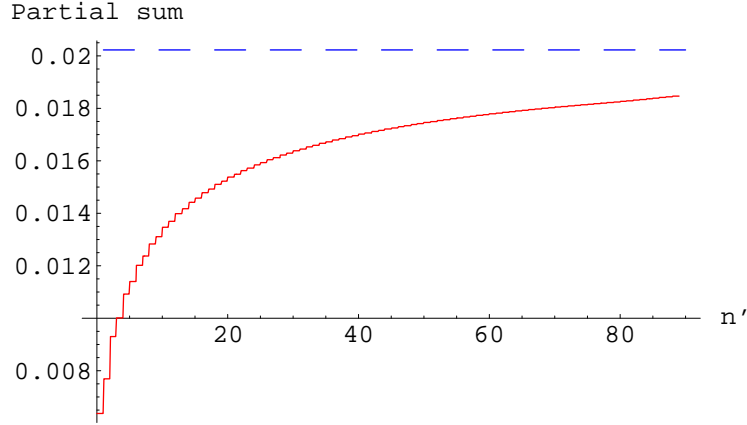}
\caption{\it Analysis of the (off-diagonal) second term in the right-hand side of 
Eq. (\ref{equality}). 
We have taken the values $n=3$ and $m_c=\beta$, $m_{s}=\beta$ and 
$m_Q=10\beta$. The dashed blue line corresponds to the difference 
between the left and right hand side of the equality if the off-diagonal term 
is neglected. The solid red line represents the contribution of the 
off-diagonal term as a function of the number of intermediate states added to the sum. 
For n' larger than 80 the numerical stability of the computation is doubtful.\label{n3}}
\end{center}
 \vspace{1mm}
\end{figure}

We could also do the 
computation with the kinematics $q^+=0$, $q^-=xM_{H_Q}$, along the lines of Ref. \cite{Bigi:1998kc}. 
This is the "spatial" component of the momentum. In this case we would obtain 
($P_{H_Q}^+=P_{H_Q}^-=M_{H_Q}
=P_n^+$ 
and $x=1-P_n^-/P_{H_Q}^-=1-M_n^2/M_{H_Q}^2$)
\be
\label{dGdx-}
\frac{d\Gamma^{(-)}}{dx}=\frac{G^2}{32\pi}\sum_{M_{n}\le M_{H_Q}}
x\Big| \langle n;P_n^+|\bar{\psi}_c(0) \gamma^+ Q(0)|H_Q \rangle\Big|^2
\delta\lp P_{H_Q}^--P_n^--q^-\rp
\,.
\ee
Note that in this case we have to compute the matrix elements in the limit $q^2, 
P_{H_Q}^+-P_n^+
\rightarrow 0$, which considerably simplifies the computation and one obtains
\be
\frac{d\Gamma^{(-)}}{dx}=\sum_{M_{n}\le M_{H_Q}} \frac{\Gamma_n}{2} \delta\left(x-1+\frac{M_n^2}{M_{H_Q}^2} \right)
\ee
for the differential decay rate, where $\Gamma_n$ is 
\be
\label{Gn}
\Gamma_n=
\frac{G^2}{4\pi}\frac{M_{H_Q}^2-M_n^2}{M_{H_Q}}
\lc\int_0^1dz \phi^{cs}_n(z)\phi_{H_Q}(z)\rc^2
\ .
\ee  
In principle, this expression should be 
equal to Eq. (\ref{dGdx+}) for all $x$. This implies the following remarkable 
identity among matrix elements ($x_n=1-M_n^2/M_{H_Q}^2$)
\bea
\label{equality}
&&
\Bigg|\int_0^1dz \phi^{cs}_n(z)\phi_{H_Q}(z)\Bigg|
=
\Bigg|
\frac{m_Qm_{c}}{(P_{H_Q}^{+})^2}\int_0^1dz\frac{\phi_{H_Q}(x_n+(1-x_n)z)\phi^{cs}_n(z)}{(x_n+(1-x_n)z)z}
\\
\nn
&&
-\beta^2\frac{1-x_n}{(P_{H_Q}^+)^2}
\sum_{n'=0}^{\infty}
(-1)^{n'}
\int_0^1\int_0^1\int_0^1dydtdz
\frac{\phi^{Qc}_{n'}(y)\phi^{cs}_n(t)\phi^{Qc}_{n'}(z)}{(t(1-x_n)+(1-z)x_n)^2}
\\
\nn
&&
\qquad\qquad\qquad
\times
(\phi_{H_Q}(x_n+(1-x_n)t)-\phi_{H_Q}(x_nz))
\Bigg|
\,.
\eea
In the left-hand-side (LHS) of the equality only the "diagonal" 
term of the "+" current contributes. For the right-hand-side 
(RHS), the first term is the "diagonal" contribution of the "-" current 
and the second term is the "off-diagonal" one. 
Note that, in principle, we cannot fix the relative sign between both 
matrix elements.
Note also that the equality Eq. (\ref{equality}) provides different information 
than Eq. (\ref{currcons}), since it relates matrix elements with different "$x$" 
($P_{H_Q}^+-P_n^+=0$ in the LHS of the equation and $P_{H_Q}^+-P_n^+=xM_{H_Q}$ in the RHS 
of the equation).
We have not been able to find a general analytic proof of these remarkable 
identities, though we have been able to do some partial checks, either when 
we have considered moments, or by using the layer functions for the 
final state (this implicitly 
assumes that we are working with a final state with a large quantum number $n$). 
In those cases we have been able to perform a comparison within an expansion in 
$1/m_Q$ and check the low order terms in this expansion. Irrespectively of the 
above, we have been able to check the equality (\ref{equality}) numerically to 
a level below the 1 \% using the numerical solution to the 't Hooft equation 
obtained from the Brower-Spence-Weis 
improvement of the Multhopp technique \cite{Brower:1978wm}. We have used two set 
of values for the masses of the quarks: $m_Q=15\beta$, $m_c=10\beta$, $m_s=0.56\beta$ 
and $m_Q=10\beta$, $m_c=\beta$, $m_s=\beta$.  
We show the comparison in Tables \ref{matrixelements} and \ref{matrixelements2}. This agreement 
is quite remarkable 
if we take into account that the support functions in both integrals are 
quite different (see Fig. \ref{supportplots}), specially for the second set 
of parameters. Note that we do not consider the second term of the RHS 
of Eq. (\ref{equality}) in the plot. 
This term appears to be a correction compared with the first one 
and vanishes in the limit $m_Q \rightarrow \infty$. This is illustrated in Tables 
\ref{matrixelements} and \ref{matrixelements2}, where this second 
term appears to be smaller for larger values of the heavy quark mass. 
This points to the fact that their scaling may go like $\sim 
\beta^2/m_Q^2$ and that there are no terms of the type $\sim 
m_c^2/m_Q^2$. The convergence of this second term is very slow. 
One has to sum over a 
very large number of states to converge to the final value. We illustrate 
this problem in Fig. \ref{n3}, where the sum is over 100 states.  

\begin{table}
\begin{center}
\begin{tabular}{|c|c|c|c|c|c|}
\hline
n &  LHS   &   RHS ("diag" term)   &  Rel. Err.  & RHS   &  Rel. Err. \\
\hline
0 & 0.96433    & 0.97144      &  $7~10^{-3}$    & 0.96627   & $2~10^{-3}$  \\
1 & 0.25999    & 0.26175      &  $6~10^{-3}$    & 0.26025   & $1~10^{-3}$  \\
2 & 0.04432    & 0.04473      &  $9~10^{-3}$    & 0.04445   & $3~10^{-3}$  \\
3 & 0.02389    & 0.02405      &  $7~10^{-3}$    & 0.02391   & $1~10^{-3}$  \\
4 & 0.00543    & 0.00538      &  $7~10^{-3}$    & 0.00540   & $3~10^{-3}$  \\
\hline
\end{tabular}
\end{center}
\caption{ \it Values for the matrix elements as defined in Eq. (\ref{equality}). 
The first column corresponds to the principal quantum number.
The second column corresponds to the left-hand side of the equality. 
The third column corresponds to the first term of right-hand side of the equality 
(the "diagonal" term). The fourth column to the relative difference between the second 
and third column. The fifth column corresponds to the right-hand side of Eq. (\ref{equality})
and the last column to the relative difference between the left and right-hand side of Eq. 
(\ref{equality}). In order 
to ease the comparison with the results of Lebed and Uraltsev \cite{lebed},  
we take $m_Q=15\beta$, $m_c=10\beta$ and $m_{s}=0.56\beta$.}
\label{matrixelements}
\end{table}

\begin{table}
\begin{center}
\begin{tabular}{|c|c|c|c|c|c|}
\hline
n &  LHS   &  RHS ("diag" term)   &  Rel. Err.  &   RHS &  Rel. Err. \\
\hline
0    &   0.46946    &  0.42992     &  $9~10^{-2}$  & 0.46493  & $8~10^{-3}$ \\
1 &      0.61406    &  0.62957     & $2~10^{-2}$   & 0.61537   &$2~10^{-3}$  \\
2 &      0.49594    &  0.51617     & $4~10^{-2}$   & 0.49821   &$3~10^{-3}$ \\
3 &      0.32820    &  0.34773     & $6~10^{-2}$   & 0.32985   & $5~10^{-3}$  \\
4 &      0.18571    &  0.19712     & $6~10^{-2}$   & 0.18694  & $6~10^{-3}$  \\
\hline
\end{tabular}
\end{center}
\caption{\it As in Table \ref{matrixelements} with the values
 $m_Q=10\beta$, $m_{c}=\beta$, $m_{s}=\beta$.}
\label{matrixelements2}
\end{table}

We are also able to 
compare with the numerical evaluation of the matrix elements (the 
LHS of Eq. (\ref{equality})) performed in Ref. \cite{lebed}. We have 
checked that our results agree with theirs within the expected numerical 
uncertainties. 

We would also like to remark that the RHS of Eq. (\ref{equality}) can be understood 
as a function of $x$, which for $x=x_n$ is equal to the LHS. Therefore, it provides 
with a definition of a continuous function in $x$. This will be relevant later on 
when trying to connect with computations using effective field theories.

So far these expressions are exact. At this stage we can perform an expansion in $1/m_Q$ and consider the 
large $n$ limit (therefore our result will hold for the OPE or SCETI region but not for the SCETII 
kinematical situation). At lowest order in those expansions, 
and using the properties of the layer function defined in sec. \ref{secthooft},
we obtain for the "diagonal" term of the matrix element (we also include the subleading corrections in $m_c/m_Q$, which can also 
be reliably computed with the layer function)
\bea
\label{eqlayer}
&&
\frac{m_Qm_{c}}{M_{H_Q}^2}\int_0^1dz\frac{\phi_{H_Q}(x+(1-x)z)\phi^{cs}_n(z)}{(x+(1-x)z)z}
=
\frac{m_Q}{M_{H_Q}^2}
\int_x^1 dy\frac{\phi_{H_Q}(y)}{y}\frac{m_c}{y-x}\phi^{cs}_n\lp\frac{y-x}{1-x}\rp
\\
\nn
&&
\simeq
\frac{m_Qm_{c}}{M_{H_Q}^2}
\left(
\frac{\phi_{H_Q}(x)}{x}\int_0^{\infty}\frac{d\xi}{\xi}{\phi}_{c}(\xi)
+\phi'_{H_Q}(x)\frac{1}{M_n^2}\frac{1-x}{x}\int_0^{\infty}d\xi {\phi}_{c}(\xi)
\right.
\\
\nn
&&
\left.
\qquad\qquad
-\frac{\phi_{H_Q}(x)}{x^2}\frac{1-x}{M_n^2}\int_0^{\infty}d\xi{\phi}_{c}(\xi)
\right)
+\cdots
\\
\nn
&&=\frac{m_Q}{M_{H_Q}^2}
\frac{\phi_{H_Q}(x)}{x}\pi\beta
\left(1+\frac{m_{c}^2}{M_{H_Q}^2}\frac{\phi'_{H_Q}(x)}{\phi_{H_Q}(x)}
-\frac{1}{x}\frac{m_{c}^2}{M_{H_Q}^2}
+\cdots\right)
\eea
by defining
\be 
\frac{y-x}{1-x}=\frac{\xi}{M_n^2}
\ee 
and expanding in $1-x$. 

\begin{figure}[h!!]
\begin{center}
\epsfxsize=8truecm \epsfbox{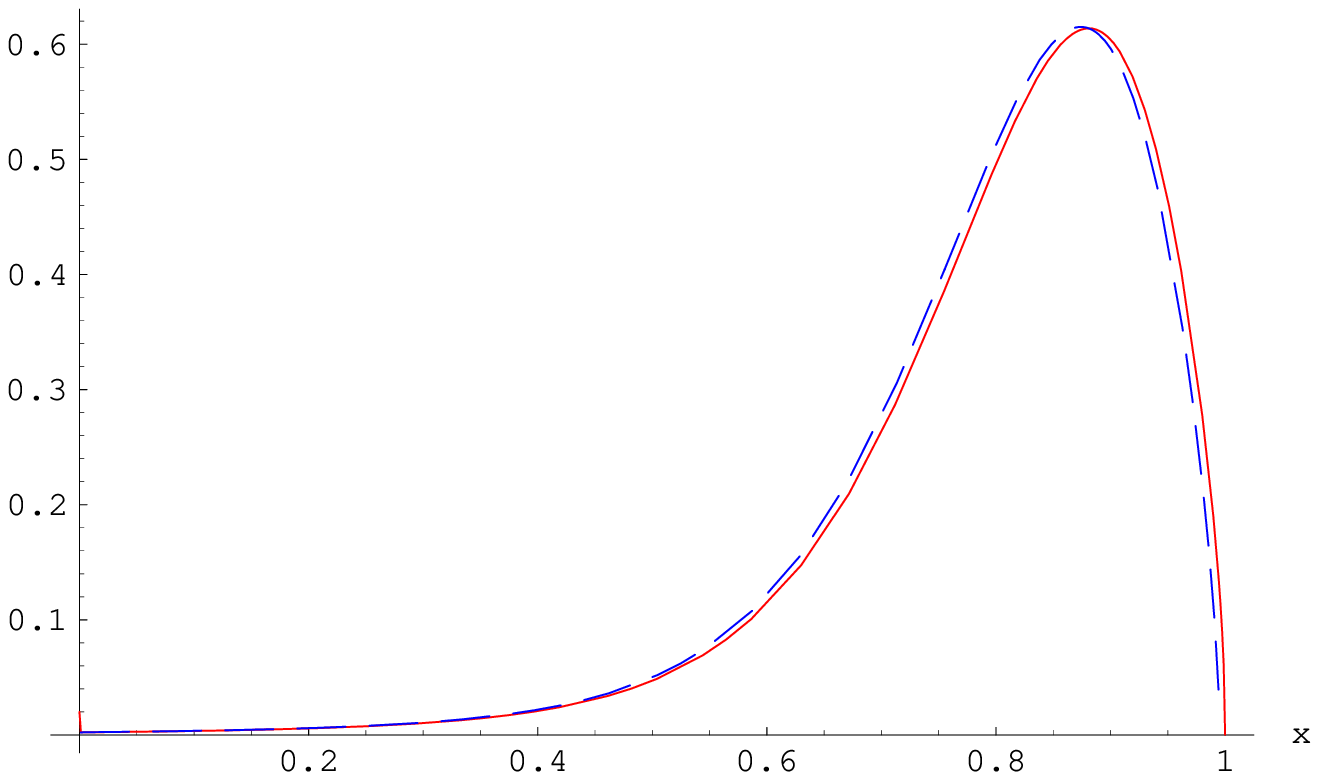}
\epsfxsize=8truecm \epsfbox{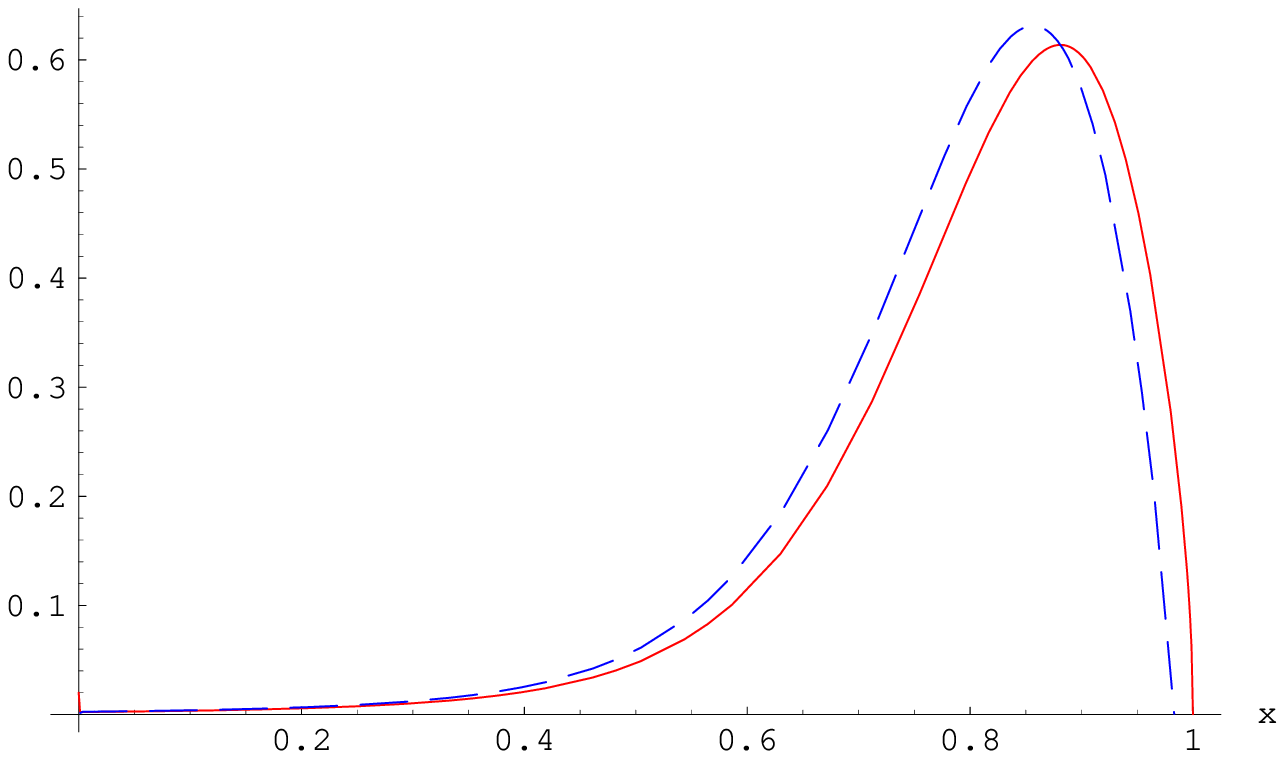}
\epsfxsize=8truecm \epsfbox{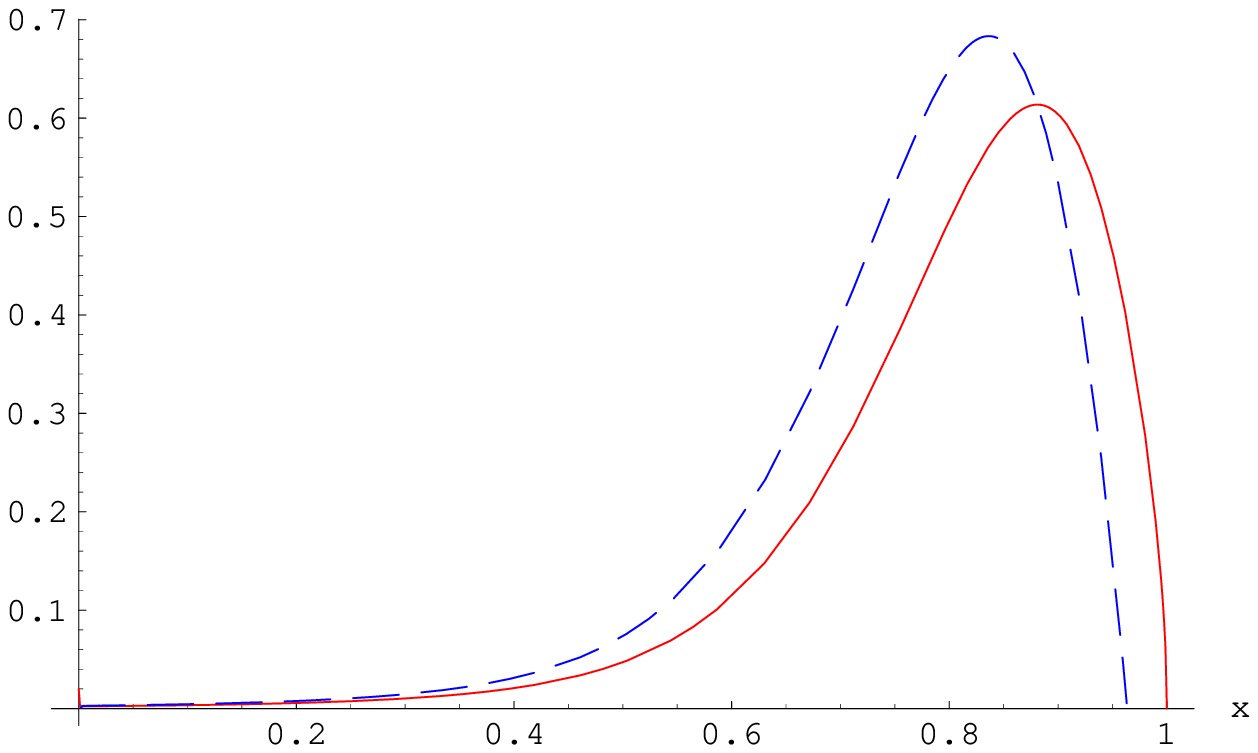}
\caption{\label{fig7}\it 
The solid red line is the LO expression of the 
RHS of Eq. (\ref{eqlayer}). The dashed blue line is 
the NLO result. We take $m_Q=10\beta$ and $m_s=\beta$, 
whereas $m_c=\beta$, $2\beta$ and $3\beta$ for the 
first, second, and third figures, respectively.}
\end{center}
 \vspace{1mm}
\end{figure}

We would like to stress that both the LHS and RHS of 
Eq. (\ref{eqlayer}) can be defined for any $x$ and not only for $x=x_n$. 
Nevertheless, they correspond to the physical matrix element only for 
$x=x_n$. On the other hand the RHS of Eq. (\ref{eqlayer}) provides 
with an interpolating function for the matrix elements at different $x_n$, 
which is independent of the dynamics of the final state $\phi_n$. 
We plot this function in Fig. \ref{fig7}. We can see that the 
next-to-leading order (NLO) is a correction compared with the leading 
order (LO), for all values of $x$ and the hard-collinear mass we consider. 
For instance, if we take $x=0.5$, we obtain $0.050022=0.0470693+0.0029527$, 
where the first term is the LO result and the second one the NLO correction. 
It should be noticed that most of the contribution 
to the NLO result comes from the derivative term of the wave-function. We will 
discuss further this issue in sec. \ref{momhadr}, when we consider the moments.   

\begin{figure}[h!!]
\begin{center}
\makebox[1.0cm]{\phantom b}
\epsfxsize=8truecm \epsfbox{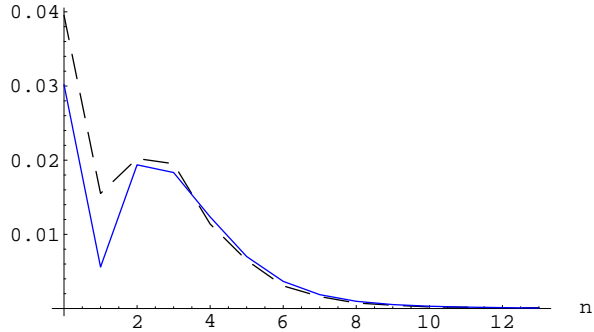}
\caption{\label{checkoffdiag}\it Plot of the absolute value of the LHS (dashed line) 
and RHS (solid line) of Eq. (\ref{offdiaglayer}) for $x=x_n=1-M_n^2/M_{H_Q}^2$. 
To improve the numerical accuracy, we use the difference between the 
LHS and the first term of the RHS of Eq. (\ref{equality}) for the numerical value of the LHS of 
Eq. (\ref{offdiaglayer}).
We take the values $m_Q=10\beta$, $m_{c}=\beta$ and $m_{s}=\beta$.}
\end{center}
 \vspace{1mm}
\end{figure}

So far we have only considered the "diagonal" term of the hadronic matrix element. For 
the "off-diagonal" term, we are, at present, not able to give approximated analytic 
results, even in the large $n$ and large $m_Q$ limit. Nevertheless, we have some 
hints about its analytic form in those limits. They come from two sources: a) the expressions 
for the moments from the hadronic computation which are already available, for $N=0$, 1 and 2, 
within an expansion in $1/m_Q$ \cite{Bigi:1998kc}; and b) the results from the effective theory 
with one-loop accuracy. As we will see, to get agreement with these results, the leading term 
in the large $n$ and large $m_Q$ limit of the "off-diagonal" correction should renormalize the 
masses of the hard-collinear and heavy quark of the "diagonal" term. In practice, one 
should have 
(although the absolute sign cannot be obtained, the relative sign with respect the "diagonal" 
term is fixed):
\bea
&&
-\beta^2\frac{1-x}{(P_{H_Q}^+)^2}
\sum_{n'=0}^{\infty}
(-1)^{n'}
\nn
\int_0^1\int_0^1\int_0^1dydtdz
\frac{\phi^{Qc}_{n'}(y)\phi^{cs}_n(t)\phi^{Qc}_{n'}(z)}{(t(1-x)+(1-z)x)^2}
\\
\nn
&&
\times
(\phi_{H_Q}(x+(1-x)t)-\phi_{H_Q}(xz))
\simeq
\frac{m_Q}{M_{H_Q}^2}
\frac{\phi_{H_Q}(x)}{x}\pi\beta
\\
&&
\qquad\qquad
\times
\left(
-\frac{\beta^2}{M_{H_Q}^2}\frac{\phi'_{H_Q}(x)}{\phi_{H_Q}(x)}
+\frac{1}{x}\frac{\beta^2}{M_{H_Q}^2}-\frac{\beta^2}{2m_{Q}^2}
+\cdots\right)
\label{offdiaglayer}
\,.
\eea
Although we were not able to check this equation on an analytic basis, we have been 
able to check it on a numerical basis. We show this comparison in Fig. \ref{checkoffdiag}. 
We can see that the LHS and RHS of Eq. (\ref{offdiaglayer}) converge to the same value as expected. 
We have also checked that if we vary the masses\footnote{One should note though that 
the computations of the 't Hooft wave function with tachyonic masses are problematic at the numerical level. This 
problem affects the accuracy of the numerical results and is more acute if one consider the derivate of the wave-function on a point-to-point 
basis.}, the same pattern survives (as far as 
the heavy quark mass is large enough). Note that this term is a correction in the $1/m_Q$ expansion. 

\begin{figure}[h]
\begin{center}
\includegraphics[width=0.79\columnwidth]{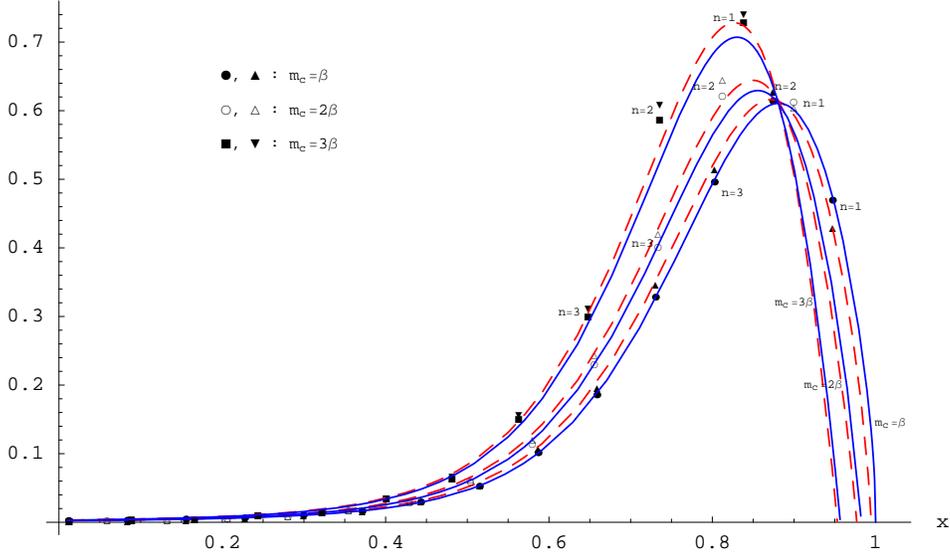}
\caption{\it The solid blue lines represent Eq. (\ref{optionBR})
for values of $m_c=\beta$, $2\beta$ and $3\beta$. The dashed red lines represent Eq. (\ref{optionBR})
for values of $m_c=\beta$, $2\beta$ and $3\beta$ replacing all the renormalized masses by their bare 
values:  $m_{i,R} \rightarrow m_i$. The dots or squares represent the LHS of Eq. (\ref{equality})
for different values of $n$. The triangles represent the diagonal term of the RHS of Eq. 
(\ref{equality}). In both cases the  values of $m_c$ are $\beta$, $2\beta$ and $3\beta$. 
We take $m_Q=10\beta$ and $m_s=\beta$ in all cases.} 
\label{difrate}
\end{center}
\end{figure}

\begin{figure}[h]
\begin{center}
\includegraphics[width=0.79\columnwidth]{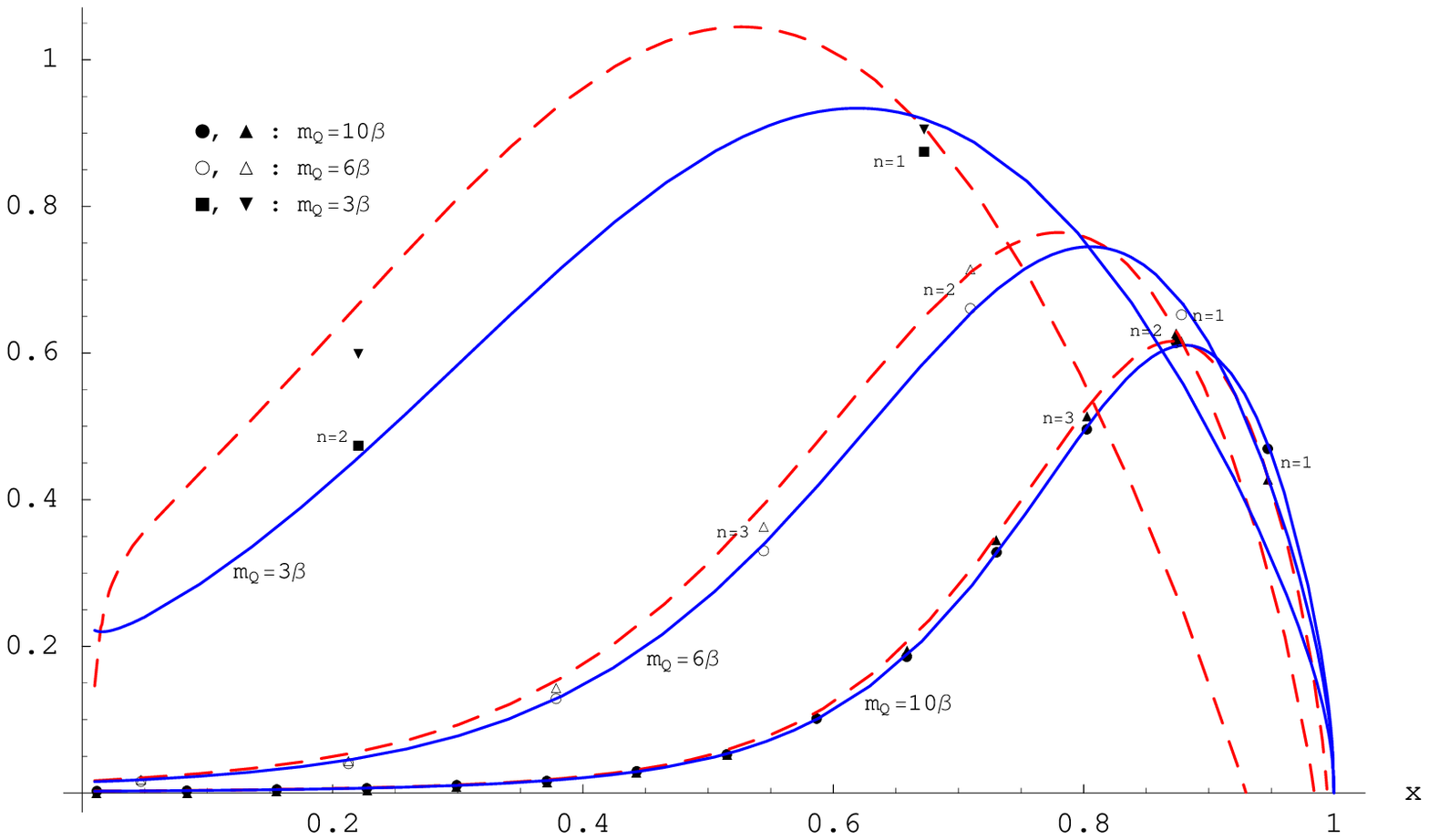}
\caption{\it The solid blue lines represent Eq. (\ref{optionBR})
for values of $m_Q=10\beta$, $6\beta$ and $3\beta$. The dashed red lines represent Eq. (\ref{optionBR})
for values of $m_Q=10\beta$, $6\beta$ and $3\beta$ replacing all the renormalized masses by their bare 
values:  $m_{i,R} \rightarrow m_i$. The squares or dots represent The LHS of Eq. (\ref{equality})
for different values of $n$. The triangles represent the diagonal term of the RHS of Eq. 
(\ref{equality}). 
We take $m_c=\beta$ and $m_s=\beta$ in all cases.} 
\label{difratemQ}
\end{center}
\end{figure}

Overall, we find that the total="diagonal"+"off-diagonal" matrix element can be written 
in the following way (up to a global sign) for large $n$ and $m_Q$:
\be
\label{optionBR}
\int_0^1dz \phi^{cs}_n(z)\phi_{H_Q}(z)
\simeq
\pi\beta\frac{m_{Q,R}}{M_{H_Q}^2}
\frac{1}{x}
\phi_{H_Q}\lp x\rp \lp 1+\frac{m_{c,R}^2}{m_{Q,R}^2}
\left(
\frac{\phi^{\prime}_{H_Q}\lp x\rp}{\phi_{H_Q}\lp x\rp}-\frac{1}{x}
\rp\rp
\Bigg|_{x=x_n} 
\,.
\ee
Note that the RHS of Eq. (\ref{optionBR}) can be understood as a function of $x$. 
Strictly speaking this expression is singular for $x \rightarrow 0$. Nevertheless, 
this effect only shows up for very small values of $x$, which are not included in the 
plots. 
We can check how well this curve compares with the exact hadronic computation 
for different values of the hard-collinear and heavy quark masses. We show the 
results in Figs. \ref{difrate} and \ref{difratemQ}. 
On the one hand we plot the hadronic matrix elements: 
The LHS of Eq. (\ref{equality}) and the diagonal term of the RHS of Eq. 
(\ref{equality}). This allows to visualize the difference of working with renormalized and 
not-renormalized masses (actually the only place where this difference is 
visible is in the hard-collinear mass multiplying the derivative of the 
't Hooft wave function of the $H_Q$ meson). On the other hand we plot
the function obtained from the boundary-layer approximation, Eq. (\ref{optionBR}), for 
all values of $x$, also with renormalized and 
not-renormalized masses. 
We can see that the agreement between the 
layer-function and the hadronic result is very good up to very low values of $n$ or, 
in other words, up to quite near the $x \rightarrow 1$ limit\footnote{
Actually, the agreement is even too good for the $x \rightarrow 1$ limit, where the 
layer-function approximation, in principle, does not apply. One could not ruled out this 
to be a numerical accident. For instance, if we use 
\be
\label{optionAR}
\int_0^1dz \phi^{cs}_n(z)\phi_{H_Q}(z)
\simeq
\pi\beta\frac{m_{Q,R}}{M_{H_Q}^2}
\frac{1}{x+\frac{m_{c,R}^2}{m_{Q,R}^2}}
\phi_{H_Q}\lp x+\frac{m_{c,R}^2}{m_{Q,R}^2}\rp
\Bigg|_{x=x_n} 
\,,
\ee
instead of Eq. (\ref{optionBR}), which is also correct to the order or interest,  
the agreement is less good in the $x \rightarrow 1$ region. On the other hand, 
Eq. (\ref{optionAR}) incorporates subleading partial effects which may jeopardize 
the agreement.}. 
Our results are also quite good up to relatively low values of the heavy quark mass. 
We can also see that the dependence on the hard-collinear mass is very 
well understood with our analytic formula. Moreover, we can also see how the 
effect of the non-diagonal term of the RHS of Eq. (\ref{equality}) is 
equivalent to renormalizing the heavy quark and hard-collinear mass in the 
region where we can trust our results\footnote{In the effective field theory, 
this renormalization would be produced by one-loop ($\sim \beta^2$) corrections. 
We can see that their effects are very tiny for basically all values of $x$. 
In the hadronic computation it reflects in the 
fact that the "off-diagonal" effects are also very small. Actually, the basic 
effect that it is seen is the renormalization of the hard-collinear mass that 
appears in the hard-collinear propagator in the effective theory. This is what is to be expected in the SCETI region, 
in the OPE region the $1/m_Q^2$ corrections are too small to be seen by the eye 
(unless the hard-collinear mass is large enough).}. Overall, we get a very consistent picture. 


We find the equality Eq. (\ref{optionBR}) quite remarkable. It implies 
that the partial decay width $\Gamma_n$ becomes independent of the final state 
wave function properties for higher excitations. 
The dependence on the final state only 
appears through $x_n=1-M_n^2/M_{H_Q}^2$:
\be
\label{Gnlayer}
\Gamma_n\stackrel{n\rightarrow\infty}{=}
\frac{G^2M_{H_Q}}{4\pi}
\frac{m_{Q,R}^2}{M^2_{H_Q}}
\frac{\pi^2 \beta^2}{M^2_{H_Q}} 
\frac{1}{x_n}
\phi_{H_Q}^2\left(
x_n
\right)
\left[1+2\frac{m_{c,R}^2}{m_{Q}^2}
\left(
\frac{\phi_{H_Q}^{\prime}(x_n)}{\phi_{H_Q}(x_n)}
-\frac{1}{x_n}\right)
+\cdots
\right]
\,.
\ee

The differential decay rate then reads
\bea
\label{dgammalayer}
&&
\frac{d\Gamma^{(+)}}{dx}=\frac{1}{2}\sum_{M_n \leq M_{H_Q}} 
\frac{G^2M_{H_Q}}{4\pi}
\frac{m_{Q,R}^2}{M^2_{H_Q}}
\frac{\pi^2 \beta^2}{M^2_{H_Q}} 
\frac{1}{x_n}
\phi_{H_Q}^2(x_n)
\\
\nn
&&
\times
\left[1+2\frac{m_{c,R}^2}{m_{Q}^2}
\lp
\frac{\phi_{H_Q}^{\prime}(x_n)}{\phi_{H_Q}(x_n)}
-\frac{1}{x_n}
\rp
+\cdots\right]
\delta\lp x-1+\frac{M_n^2}{M_{H_Q}^2}\rp
\,.
\eea
This expression will be suitable to a smoother connection with the computation 
using effective field theories. 

Eq. (\ref{dgammalayer}) applies to the kinematical situation $ 1-x \gg \beta^2/m_Q^2$. 
It includes the leading term in an expansion in $1/m_Q$ and 
$1/n$. Some kinematical $1/m_Q$ corrections are automatically included 
by working with the exact $\phi_{H_Q}$ wave function instead of working 
with the strict static limit. The corrections of order $m_c^2/m_Q^2$, 
$\beta^2/m_Q^2$ have also been included. 
In principle, this expression could be systematically improved by 
considering corrections in $1/m_Q$ and $1/n$. Nevertheless, this would require 
to know the corrections in $1/n$ to the integrals that appear in our expressions, 
which at present are not known. We expect to study 
further this issue in the future. 

Finally, we would like to stress that the limit $m_c \rightarrow 0$  has 
to be taken with care, as it is evident from Eq. (\ref{equality}). A naive limit 
$m_c \rightarrow 0$ may lead to wrong results.

\subsection{Moments}
\label{momhadr}
The differential decay rate is not a very well defined object in the large $N_c$, since it 
becomes either infinity or zero. In particular its comparison with the expressions 
obtained from effective theories that use perturbative factorization is 
not possible, as we will see in the next sections: on the one hand one obtains 
a series of delta terms, whereas on the other 
one gets an smooth function of $x$\footnote{
Real experimental data on semileptonic B meson decays is
usually available in terms of moments, and therefore
so are the corresponding theoretical predictions.
In Ref. \cite{Rojo:2006kr} the differential decay rate itself
was reconstructed from available experimental information
on its moments, allowing thus for a more general comparison between
theory and experiment.}. 
At this respect one may think that it is better to work with moments\footnote{
This actually does not cause the problem of quark-hadron 
duality to vanish, though, as it has been emphasized in Ref. \cite{Bigi:1998kc} for the inclusive decay width.}:
\be
M_N \equiv \int_0^1 dx x^{N-1} \frac{d\Gamma}{dx}
\,.
\ee

Exact expressions for the moments in terms of hadronic matrix elements 
can be obtained by using the expressions for 
$\frac{d\Gamma^{(+)}}{dx}$ or $\frac{d\Gamma^{(-)}}{dx}$ obtained in the previous 
subsection:
\bea
\label{MNhadrnos}
M_N&=&
\frac{G^2M_{H_Q}}{4\pi}\sum_{M_{n}\le M_{H_Q}}
x_n^{N}
\Bigg|
\frac{m_Qm_{c}}{(P_{H_Q}^{+})^2}\int_0^1dz\frac{\phi_{H_Q}(x_n+(1-x_n)z)\phi^{cs}_n(z)}{(x_n+(1-x_n)z)z}
\\
\nn
&&
-\beta^2\frac{1-x_n}{(P_{H_Q}^+)^2}
\sum_{n'=0}^{\infty}
(-1)^{n'}
\int_0^1\int_0^1\int_0^1dydtdz
\frac{\phi^{Qc}_{n'}(y)\phi^{cs}_n(t)\phi^{Qc}_{n'}(z)}{(t(1-x_n)+(1-z)x_n)^2}
\\
\nn
&&
\qquad\qquad\qquad
\times
(\phi_{H_Q}(x_n+(1-x_n)t)-\phi_{H_Q}(x_nz))
\Bigg|^2
\,,
\eea 
\be
\label{MNhadrBigi}
M_N=
\frac{G^2M_{H_Q}}{4\pi}\sum_{M_{n}\le M_{H_Q}}
x_n^{N}
\Big| \int_0^1 dy\phi^{cs}_n(y)\phi_{H_Q}(y)\Big|^2
\,,
\ee 
where $x_n=1-M_n^2/M_{H_Q}^2$. As we have mentioned previously both 
expressions for the moments yield the same result. 

In order to perform some analytical analysis, it is necessary to 
be able to compute the different matrix elements or, at least, the sum of 
matrix elements that contributes to the moments. In general this is not possible. 
Nevertheless, for 
some specific cases, it is possible to obtain approximated expressions. If 
we take $M_1$ from Eq. (\ref{MNhadrBigi}), it corresponds to the total 
decay width. In this case 
it is possible \cite{Bigi:1998kc} to obtain a closed analytic 
expression up to $O(1/M_{H_Q}^5)$ suppressed corrections in terms 
of expectation values of matrix elements of the $H_Q$-meson wave function 
 by using sum rules:
\bea
\label{M1bigi}
M_1\equiv \Gamma_{H_Q}&=&\frac{G^2}{4\pi} \frac{(m_Q^2-m_c^2)}{M_{H_Q}}
\int_0^1 \frac{dx}{x}\phi^2_{H_Q}(x)-\sum_{M_n\ge M_{H_Q}} \Gamma_n
\\
&=&\Gamma_Q\left[\frac{m_Q}{M_{H_Q}}\int_0^1\frac{dx}{x}\phi_{H_Q}^2(x)
+
O\lp\frac{1}{m_Q^5}\rp\right]
\,,
\eea
where
\be
\Gamma_Q=\frac{G^2}{4\pi}\frac{m_Q^2-m_c^2}{m_Q}
\ee
is the free heavy quark decay rate, 
and $\Gamma_n$ has been defined in Eq. (\ref{Gn}) (note that in 
Eq. (\ref{M1bigi}) 
they represent partial decay widths that are not allowed by momentum conservation).

For $N \not= 0$, in general, it is not possible to follow the same procedure, 
since the sum rules become divergent. Only for $N=0,2$, it is also possible to 
obtain a finite result:
\be
M_0=\frac{G^2M_{H_Q}}{4\pi}\left(1+O\left(\frac{1}{m_Q^5}\right)\right)
\,,
\ee
\bea
\nn
M_2&=&\frac{G^2M_{H_Q}}{4\pi} \frac{(m_Q^2-m_c^2)^2}{M_{H_Q}^4}
\int_0^1 \frac{dy}{y^2}\phi^2_{H_Q}(y)-\frac{1}{M_{H_Q}^2}
\sum_{M_n\ge M_{H_Q}} (M_{H_Q}^2-M_n^2)\Gamma_n
\\
&=&\Gamma_Q\left[\frac{m_Q}{M_{H_Q}}\frac{m_Q^2-m_c^2}{M_{H_Q}^2}
\int_0^1\frac{dx}{x^2}\phi_{H_Q}^2(x)
+
O\lp\frac{1}{m_Q^3}\rp\right]
\,.
\eea
The $N=0$ case is basically due to probability conservation. It can be noticed that 
the above moments can be written in a more compact way in the following form:
\be
\label{MNbigi}
M_N=\frac{G^2M_{H_Q}}{4\pi}
\left[
\left(\frac{m_Q^2-m_c^2}{M_{H_Q}^2}\right)^N
\int_0^1\frac{dx}{x^N}\phi_{H_Q}^2(x)
+
O\lp\frac{1}{m_Q^3}\rp
\right], \qquad {\rm for} \; N=0,1,2.
\ee
For $N$ larger than two the integral becomes divergent.

The above expressions contain some implicit dependence on the heavy quark mass, 
since, so far, we have used the exact $H_Q$-meson wave function. If we perform 
an explicit expansion in $1/m_Q$, one obtains, up to ${O}(1/m_Q^3)$,
\be
M_0=\frac{G^2m_{Q}}{4\pi}
\left[1+\frac{\langle t \rangle}{m_Q}-\frac{\langle t\rangle^2-3 \langle t^2\rangle+\beta^2}{2m_Q^2}+{O}\left(\frac{1}{m_Q^3}\right)\right]
\,,
\ee
\be
M_1=
\frac{G^2m_{Q}}{4\pi}
\left[1+\frac{ \langle t\rangle^2- \langle t^2\rangle+\beta^2-2m_c^2}{2m_Q^2}+{O}\left(\frac{1}{m_Q^3}\right)\right]
\,,
\ee 
\be
M_2=
\frac{G^2m_{Q}}{4\pi}
\left[1-\frac{\langle t\rangle}{m_Q}+\frac{3\langle t\rangle^2-3\langle t^2\rangle+3\beta^2-4m_c^2}{2m_Q^2}+{O}\left(\frac{1}{m_Q^3}\right)\right]
\,,
\ee
where the static limit expectation values are defined 
in Eq. (\ref{tavn}).
We relegate the numerical comparison of these expressions 
with the exact ones 
to sec. \ref{comparison}, since these expressions will also be 
obtained from  
the effective theory computation.

The above expressions for the moments have been obtained for low $N$. Therefore, 
they correspond, somewhat, to the kinematical regime where the OPE is valid. We 
may consider to use the approximated expression obtained 
using the properties of the final state wave function for large $n$, 
i.e. the layer function, for 
$\frac{d\Gamma^{(+)}}{dx}$ in Eq. (\ref{dgammalayer}). 
Therefore, we expect the expressions that we will obtain to be also  
valid for larger values of $N$: $ N \le M_Q/\beta$ ($1-x \sim \beta/m_Q$), up to corrections 
of order $N\beta^2/m_Q^2 \sim \beta^2/M_n^2$. This means the 
kinematical regime where the OPE and SCETI are valid.  
 Note, however, that we integrate for all $x$ in the moments. 
Therefore, this includes contributions from $x \sim 1$, equivalent to final states 
with $n \sim 1$ for which the layer-function approximation is not valid. 
A very rough estimate sets the contribution of these states to the 
moments (for $N \sim 1$) of $O(1/m_Q^4)$ or smaller. In any case the fact that 
we have problems to obtain approximate analytic expressions for  
Eq. (\ref{offdiaglayer}) sets the accuracy of the calculation. The expressions that we obtain for the moments read
\bea
M_N 
&\simeq&
\frac{G^2M_{H_Q}}{4\pi}
\frac{m_{Q,R}^2}{M^2_{H_Q}}
\frac{\pi^2 \beta^2}{M_{H_Q}^2}
\sum_{M_n \leq M_{H_Q}}
\int_0^1 dx x^{N}
\frac{\phi_{H_Q}^2\left(x\right)}{x}
\\
\nn
&&
\times
\left[1+2\frac{m_{c,R}^2}{m_{Q}^2}
\lp
\frac{\phi_{H_Q}^{\prime}(x)}{\phi_{H_Q}(x)}
-\frac{1}{x}
\rp
\right]
\delta\lp x-1+\frac{M_n^2}{M_{H_Q}^2}\rp
+\cdots
\\
\nn
&\simeq& 
\frac{G^2M_{H_Q}}{4\pi}
\frac{m_{Q,R}^2}{M^2_{H_Q}}
\frac{\pi^2 \beta^2}{M_{H_Q}^2}
\sum_{M_n^2 \leq M^2_{H_Q}}
\int_0^1 dx x^{N}
\frac{\phi_{H_Q}^2\left(x+\frac{m_{c,R}^2}{m_{Q,R}^2}\right)}{\lp x+\frac{m_{c,R}^2}{m_{Q,R}^2}\rp^2}
\delta \lp x-1+\frac{M_n^2}{M_{H_Q}^2}\rp
+\cdots
\,,
\eea
where in the second equality we have reshuffled the NLO correction to the layer function
in a way that is correct at the accuracy of the calculation and that it 
will ease some intermediate analytic computations, making them more compact. 

By working with 
moments, which imply an integral over all $x$, it becomes possible to perform a 
Euler-McLaurin expansion for the sum over $n$, which, at lowest order, it is just 
equivalent to replace the sum by an integral: $\sum_n \rightarrow \int dn$. 
This replacement allows us to make quantitative the comparison between the perturbative 
and hadronic result. 
Note however that the Euler-McLaurin expansion is an asymptotic expansion. 
Therefore, it is a difficult question to assign an error. 
Here we will not dwell further on the error used by replacing the sum by the 
integral. To go beyond this approximation would require a better knowledge of the properties of the 
layer function and a systematic procedure to get corrections from it, which 
is relegated for future work. In any case, the result we obtain after the smearing reads
\bea
\label{MNlayersmeared}
&&M_N
=
\frac{G^2M_{H_Q}}{4\pi}
\frac{m_{Q,R}^2}{M^2_{H_Q}}
\int_0^1 dx x^{N}
\frac{\phi_{H_Q}^2\left(x+\frac{m_{c,R}^2}{m_{Q,R}^2}\right)}{\lp x+\frac{m_{c,R}^2}{m_{Q,R}^2}\rp^2}
=
\frac{G^2M_{H_Q}}{4\pi}\frac{m_{Q,R}^2}{M_{H_Q}^2}
\\
\nn
&&
\times
\int_0^1
\lp1-\frac{m_{c,R}^2}{xm_{Q,R}^2}\rp^N  
 \frac{dx}{x^2} x^{N}\phi_{H_Q}^2\lp x\rp
\simeq 
\frac{G^2M_{H_Q}}{4\pi}\frac{m_{Q,R}^2}{M_{H_Q}^2}
\lp 1-\frac{m_{c,R}^2}{m_{Q,R}^2}\rp^N 
\int_0^1 
 \frac{dx}{x^2} x^{N}\phi_{H_Q}^2\lp x\rp
 \,,
\eea
where in the last equality we have used $\frac{m_{c,R}^2}{xm_{Q,R}^2} \simeq 
\frac{m_{c,R}^2}{m_{Q,R}^2}$, which is correct with the accuracy of our calculation. 
Eq. (\ref{MNlayersmeared}) is correct at leading order in the OPE and SCETI kinematic region. 
In the OPE region is correct up to, and including, $O(1/m_Q^2)$ corrections in the 
situation when it is possible to compare with the already known hadronic results 
($N=0,1,2$). Note that in order to get this agreement it is crucial to "renormalize" the 
masses of the hard-collinear and heavy quark. This renormalization effect can be traced back to 
the "off-diagonal" contribution to the "-" current.    
We can also give expressions for a general $N$
within an expansion in $1/m_Q$. At $O(1/m_Q^2)$ 
we find the following expression\footnote{Quite remarkable, we obtain the same 
expression if we extrapolate Eq. (\ref{MNbigi}) to values of $N$ different of 0, 1, 2, 
for which it was originally obtained, and we expand in $(1-x)$ before doing the integral.}
\bea
\label{MNlayerexp}
M_N^{OPE}&\simeq&\frac{G^2m_Q}{4\pi}\lp1-(N-1)\frac{\langle t\rangle}{m_Q} 
+\frac{(2N-1)\beta^2-2Nm_c^2}{2m_Q^2}\right. \\
& &\left. +\frac{\lp2(N-2)+3\rp\langle t\rangle^2+\lp (N-2)(N-3)-3\rp \langle t^2\rangle}{2m_Q^2}
+O\left(\frac{1}{m_Q^3}\right)\rp \nonumber
\,.
\eea 
The expression for $M_N$, Eq. (\ref{MNlayersmeared}), also applies to the SCETI region. 
As we have already mentioned, the above expression is correct at leading order in the 
$\beta^2N/m_Q^2$ expansion. Our expression also includes the subleading 
corrections of $O(\beta/m_Q)$. Formally, in this kinematical regime we could 
approximate $M_N$ by the following expression (note that $\phi_Q^2\lp x\rp$
should also be expanded in $1/m_Q$)
\bea
\label{MNlayerscetexp}
M_N^{SCETI}&\simeq&\frac{G^2M_{H_Q}}{4\pi}\frac{m_{Q}^2}{M_{H_Q}^2}  
\int_0^1 dx e^{-N(1-x)}\phi_Q^2\lp x\rp
\\
\nn
&&
\times
\left(1+2(1-x)-N\frac{(1-x)^2}{2}-N\frac{m_{c,R}^2}{m_{Q}^2}+\cdots\right)
\,,
\eea
where the terms neglected are of relative order $\beta^2/m_Q^2$, which, in 
principle, we cannot claim to have all of them. The reason is that 
we have not considered terms of the 
type $\beta^4/m_Q^4\partial^2 \phi^2_{H_Q}(x)/\partial^2 x$, which 
would contribute to the moments at NNLO in the SCETI region. Note again the necessity to 
renormalize the hard-collinear mass. This effect can be traced back to the derivative 
term in the boundary layer approximation of the "off-diagonal" term
 and can be unambiguously identified numerically. The reason this term is 
 enhanced is because one has contributions like 
 \be
 \int_0^1dxx^{N-2}\frac{\partial \phi^2_{H_Q}(x)}{\partial x} 
 =
 -(N-2)\int_0^1dxx^{N-3} \phi^2_{H_Q}(x)
\,.
\ee 

Finally, we would also like to consider another observable that is usually used in 
the study of the differential 
heavy meson decay rate. Since, in real life, many times one cannot measure 
over all the spectrum of final particles, one has to
introduce a cutoff to the inclusive measurement. Therefore, 
the following observable is usually considered:
\be
\Gamma_{H_Q}(y)\equiv \frac{1}{\Gamma_{H_Q}}\int_{1-y}^{1}
dx~ \frac{d\Gamma}{dx}(x) \ , \qquad  0\le y \le 1 
\ . 
\ee
We relegate a numerical analysis of this observable as well as of the moments to sec. \ref{comparison}.

\section{SCETI: Multimode approach}
\label{SCETI}
We want to describe now the hadronic results obtained in the previous 
section using effective field theories. The aim is to describe the 
decay of the heavy meson to a bunch (one in the large $N_c$) of hadronic particles 
with invariant moment $P_X^2=M_{H_Q}^2(1-x) \gg \lQ^2$. The usual procedure, 
this scale being much larger than $\lQ$, is to use perturbative computations. 
Actually, here lies the heart of the problem of quark-hadron 
duality, since we are working in the Minkowskian region and, therefore, 
near the mass-shell region. Nevertheless, we expect that by working 
with the 't Hooft model we may better visualize the problem.

In this section we first approach the problem adapting the 
present formulations  of effective theories for very energetic 
collinear particles \cite{bauerfact,Bauer:2001yt,Beneke1,neubert}, 
in particular of Ref. \cite{Beneke1}, 
to the two dimensional case, and we relegate an 
alternative approach to the next section. In those references, one 
attempts to explicitly obtain 
all the modes that one has in the theory from perturbation theory. 
This heavily relies in the concept of threshold expansion of Feynman 
diagrams \cite{Beneke:1997zp}. 

\begin{table}[t]
\begin{center}
\begin{tabular}{cc}
\multicolumn{2}{c}{SCETI fields}\\
\hline
\hline
Hard-collinear light quark  &  $\xi_{hc}$  \\
Hard-collinear gluon  &  $A^{\mu}_{hc}$  \\
Soft light quark & $q_s$ \\
Soft gluon & $A^{\mu}_s$   \\
\hline
\end{tabular}
\end{center}
\caption{\it Relevant fields in SCETI.}
\label{fields}
\end{table}

In this paper we will not exhaustively explore the different modes 
that may appear in the 't Hooft model. We will see that at the order 
we will work here it will be enough to work with hard-collinear and 
soft modes\footnote{Nevertheless, we believe the 't Hooft model provides 
a nice framework on which to explore which modes really appear in 
SCETI. We expect to pursue this line of research in the future.}, for 
which we set the notation in Table \ref{fields}. Our aim is 
to try to see explicitly at which point 
in this effective field theory derivation one 
approximates the hadronic result by a partonic one. 

\begin{table}[t]
\begin{center}
\begin{tabular}{ccccc}
\multicolumn{5}{c}{SCETI modes}\\
\hline
 Mode & $p^-$ & $p_{\perp}$ & $p^+$ & $p^2$ \\
\hline
\hline
 Hard & 1  & 1 & 1 & 1 \\
Hard-Collinear
  & 1 & $\lam$ & $\lam^2$ & $\lam^2$ \\
Soft
  & $\lambda^2$  & $\lambda^2$ &  $\lam^2$ & $\lam^4$ \\
\hline
\end{tabular}
\end{center}
\caption{ \it Relevant momentum configurations (modes) in SCETI.}
\label{modes}
\end{table}

First of all let us set the terminology of the different modes.
We have already mentioned that the small expansion parameters are 
$\lambda$ and $\bar \lambda$. We will formally work in the situation 
$\lambda \sim \bar \lambda \ll 1$, which corresponds to the SCETI kinematic 
region. Nevertheless, our results will also be valid in the OPE region where 
$\bar \lambda \sim 1$. 
In terms of $\lambda$, the momenta of the
different modes scale as shown in Table \ref{modes}. 
The only difference in $D=1+1$ is that
there
is no perpendicular component $p_{\perp}$ of the momentum. The
momentum is decomposed as
\be
p^{\mu}=n_+p\frac{n_-^{\mu}}{2}+ p_{\perp}+n_-p\frac{n_+^{\mu}}{2}
\,,
\ee
so that a collinear particle is defined as a particle with
large light-cone momentum in the $n_+$ direction. The fields 
that are relevant for the effective theory are shown 
in Table \ref{fields}. The field of each particle is decomposed into 
all the possible modes: $\psi=\xi_{hc}+\eta_{hc}+q_s+\cdots$. 
In practice only a few modes contribute to a given field. 
For instance, the quark $s$ can be approximated by 
its soft mode: $\psi_s\simeq q_{s,s}$. We also use the notation $p^+\equiv
n_+p$, and the same for other vector components. A field with
momentum $p$ varies in position space according to the uncertainty
principle,
\be
x^{\mu}=(x^+,x_{\perp},x^-)\sim(\frac{1}{p^-},\frac{1}{p_{\perp}},
\frac{1}{p^+})\ .
\ee
The scaling of the quark and gluon fields in $D$ spacetime dimensions
can be first naively estimated from the free quark and gluon propagators 
in position space quantized in the equal-time frame\footnote{In the light-cone 
quantization frame things are more complicated, in particular for the fermions, 
which have different free propagators.},
\be
\la 0 |T\lp
\psi(x)\bar{\psi}(y)\rp|0\ra=\int\frac{d^dp}{(2\pi)^4}\frac{i\ps}{p^2+
i\epsilon}
e^{-ip\cdot (x-y)} \ ,
\ee
\be
\la 0 |T\lp
A^{\mu}(x)A^{\nu}(y)\rp|0\ra=\int \frac{d^dp}{(2\pi)^4}\frac{i}{p^2+
i\epsilon}\lc -g^{\mu\nu}+(1-\alpha)\frac{p^{\mu}p^{\nu}}{p^2}\rc
e^{-ip\cdot (x-y)} \ ,
\ee
where the gluon propagator is obtained in a general covariant gauge.
Using these propagators, the scaling of the different fields of the
effective theory is the following:
\begin{itemize}
\item Soft light quark $\quad q_s\sim \lambda^{D-1}$
\item Soft gluon $\quad A_s\sim \lambda^{D-2}$
\item Hard-collinear gluon $\quad A^{+}_{hc}\sim
  \lambda^{(D-4)/2},\quad
 A^{\perp}_{hc}\sim
  \lambda^{(D-2)/2},\quad  A^{-}_{hc}\sim
  \lambda^{D/2}$
\item Hard-collinear quark $\quad \xi_{hc}\sim \lambda^{(D-2)/2}, \quad
\eta_{hc}\sim \lambda^{D/2}$
\end{itemize}
To obtain the scaling of the
hard-collinear quark fields, we have decomposed these fields using
projection operators
\be
\label{decom}
\psi_{hc}(x)=\xi_{hc}(x)+\eta_{hc}(x), \quad
\xi_{hc}(x)\equiv 
\Lambda_-\psi_{hc}(x),\quad
\eta_{hc}(x)\equiv 
\Lambda_+\psi_{hc}(x) \ .
\ee
One can check that for $D=4$ one recovers the usual scalings of the
effective theory. 
The scalings of the effective theory fields in $D=4$ and $D=2$ can be
seen in Table \ref{scalings2}. One can observe that in $D=2$ the
hard-collinear gluons have negative scalings. The solution to this 
problem comes by realizing that in $D=2$ the strong coupling $g$ has 
dimensions of mass and, actually, sets the scale of $\lQ$ so 
\be
g\sim \lam^2 \ ,
\ee
and we can see that the scaling of the hard collinear gluons times $g$ is positive.

\begin{table}[t]
\begin{center}
\begin{tabular}{ccc}
\multicolumn{3}{c}{Scaling of the SCETI fields}\\
\hline
  Field  &  $D=4$ [$g\sim\lam^0$]&  $D=2$ [$g\sim\lam^2$]\\
\hline
$\xi_{hc}$   & $\lam$ & $\lam^0$\\
$\eta_{hc}$  &  $\lam^2$      &  $\lam$      \\
\hline
$gA_{hc}^-$   & $\lam^0$ &  $\lam$\\
$gA_{hc}^{\perp}$   & $\lam$ & - \\
$gA_{hc}^+$   & $\lam^2$ & $\lam^3$ \\
\hline
$q_s$   & $\lam^3$ & $\lam^1$ \\
\hline
$gA_s^{\mu}$   &  $\lam^2$& $\lam^2$ \\
\hline
\end{tabular}
\end{center}
\caption{\it Scaling of the effective theory fields in $D=4$ and $D=2$
taking into account the scaling of the coupling.}
\label{scalings2}
\end{table}

It can be observed that the hard-collinear gluon field times g does not 
scale as the corresponding hard-collinear momentum but rather, it is 
suppressed by powers of $\lambda$. This will lead to the result that 
hard-collinear interactions are suppressed with respect to the 
corresponding kinetic terms.

A crucial point to fix the scaling of each term in the
effective action is to know the scaling of the integration element $d^2x$,
which depends on the fields present on each term. The different
scalings can be seen in Table \ref{integel}.

\begin{table}[t]
\begin{center}
\begin{tabular}{cc}
\multicolumn{2}{c}{Scaling of the integration element}\\
\hline
  Fields  &  $d^2x$ \\
\hline
hc & $\lam^{-2}$ \\
s & $\lam^{-4}$ \\
hc+s & $\lam^{-2} $  \\
\hline
\end{tabular}
\end{center}
\caption{\it Scaling of the integration element in the effective action.}
\label{integel}
\end{table}

We are now in the position to write the effective Lagrangian, which we do 
in the next section.

\subsection{Lagrangian and heavy-to-light current}

We first want to translate to two dimensions the standard procedure to 
obtain the effective Lagrangian. This first means to integrate out the 
$\eta_{hc}$ component of the hard-collinear 
field\footnote{This could be considered somewhat strange. If we were 
working in the light-front Hamiltonian frame with light-cone gauge $A^+=0$, 
$\eta_{hc}$ would correspond to the physical componentent, $\psi_{hc,+}$, of the 
field. Therefore, we would be integrating out the physical component of the 
field and keeping the constraint.}. We will only consider the leading order Lagrangian here. 

\begin{figure}[h]
\begin{center}
\includegraphics[width=0.79\columnwidth]{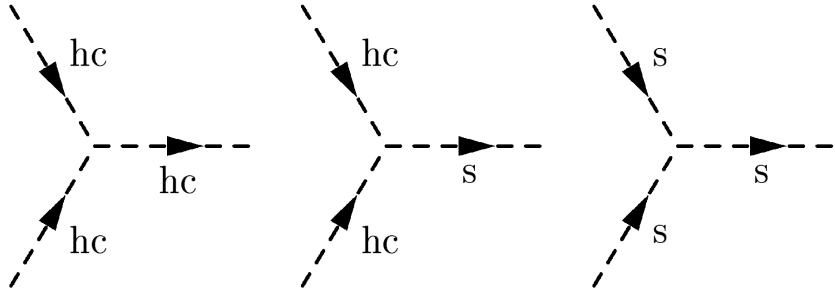}
\caption{\it Allowed vertices by momentum conservation in a theory with
  only hard-collinear and soft modes. Dashed lines can be either
quarks or gluons}
\label{vertex}
\end{center}
\end{figure}
\begin{figure}[h]
\begin{center}
\includegraphics[width=0.99\columnwidth]{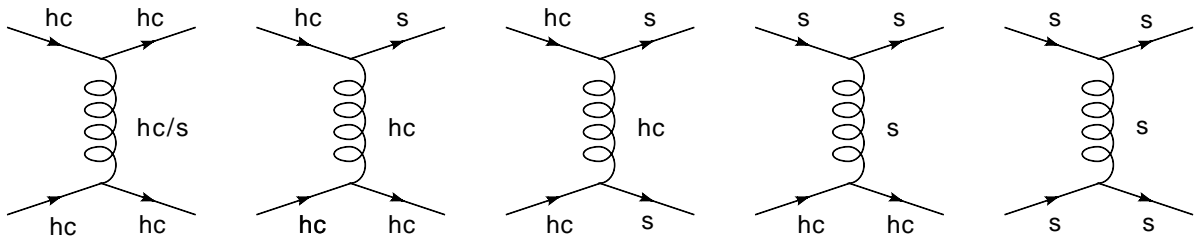}
\caption{\it Allowed scattering processes.}
\label{allowed}
\end{center}
\end{figure}

The allowed vertices by momentum conservation are drawn in Fig. \ref{vertex}. 
Since the gluons do not appear as physical particles, we are always faced 
with diagrams of the sort of those shown in Fig. \ref{allowed}. By power counting 
we can easily see that the self interactions between hard-collinears 
(in particular those including hard-collinear gluons) are suppressed 
by powers of $\lambda$. Therefore, at leading order, we only have to 
consider soft gluons and quarks and hard-collinear quarks. 
The leading order Lagrangian then reads
\be
\label{scethcslag}
\ml^{(0)}=\ml_s^{(0)}+\ml_{hc}^{(0)} \ , 
\ee
\be
\ml^{(0)}_s\equiv-\frac{1}{2}\mathrm{tr}\lc F_{\mu\nu,s}F^{\mu\nu,s}\rc+
\bar{q}_{s,s}i\Ds_s q_{s,s}+\ml_{HQET} \ ,
\ee
\be
\ml_{hc}^{(0)}=\frac{1}{4}
\mathrm{tr}\lc (D^+_S A^-_{hc}-\partial^-A^+_{hc})^2\rc
+{\bar \xi}_{c,hc}\lp in_+D_s-\frac{m_c^2-i\epsilon}
{i\partial^-}
\rp\frac{\ns_-}{2}\xi_{c,hc} \ .
\ee
In this 
expression we have kept kinematical subleading corrections 
proportional to the mass of the hard-collinear.

We do not explicitly write the HQET sector of the theory. In practice 
it will turn out more convenient to implicitly 
keep the heavy quark mass dependence 
and to expand at the end of the calculation. 

\begin{figure}[t]
   \vspace{-3cm}
   \epsfysize=27cm
   \epsfxsize=18cm
   \centerline{\epsffile{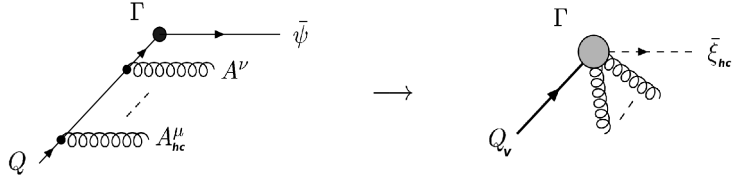}}
   \vspace*{-21cm}
\caption{\it Tree-level matching of the heavy quark current.}
\label{tran_current}
\end{figure}

The next step in order to apply SCETI to the 
semileptonic decay is to write the heavy-to-light current
$\bar{\psi}_c\Gamma Q$ (where $\Gamma=\gamma^{\mu}$ in our case) 
in terms of the effective theory fields. The emission of a 
hard-collinear quark (see Fig. \ref{tran_current}) 
by the near on-shell heavy quark puts it
off shell, and it stays off-shell when subsequent hard-collinear
and soft gluons are emitted. Therefore the effective current must
reproduce these diagrams that are absent in the effective theory, and
it can be shown that
\be
J_{QCD}=\bar{\psi}_c\Gamma Q=e^{-im_Qvx}\bar{\psi}_c\Gamma \lp
1-\frac{1}{i\Ds-m_Q(1-\vs)}g\as_{hc} \rp Q_v \ .
\ee 
We
must expand the above expression in powers of $\lam$.
Starting from
\be
\mathcal{Q}\equiv \lp
1-\frac{1}{i\Ds-m_Q(1-\vs)}g\as_{hc} \rp Q_v \ ,
\ee
and defining 
\be
S_0\equiv\frac{1}{\lc i\partial_+\frac{\ns_-}{2}-m_Q(1-\vs)\rc}=
\frac{1}{v_-}\lp\frac{\ns_-}{2m_Q}+\frac{1+\vs}{i\partial_+}\rp\sim
\lam^0 \ ,
\ee
where the scaling is like this 
because the momentum in the covariant derivative is hard-collinear,
since it is the momentum that flows through the heavy quark line
once the hard-collinear gluon has been emitted. 
Expanding in $\lam$, one arrives to the following result
\be
\mathcal{Q}=\lp 1-S_0gA_{hc+}\frac{\ns_-}{2}+S_0
gA_{hc+}\frac{\ns_-}{2}S_0gA_{hc+}\frac{\ns_-}{2}+\mo(\lam^3)\rp 
\lp 1+\mo(\lam)\rp Q_v \ .
\ee
In the last term the $\mo(\lam)$ indicates possible contributions
coming from the integration of the short component of the heavy quark field.

We would like here to stress the difference with the situation in 
4 dimensions, where a hard-collinear Wilson line appears multiplying 
the effective current. Actually here this also happens but the exponent 
is suppressed by powers of $\lambda$ (in four dimensions it is 
only suppressed by powers of $\alpha_s$).

This is the first part of the construction of the effective current. 
The second part consists on the matching of the light quark field
$\psi_c$ in terms of the effective theory fields. Since the effective
current
must be constructed such that it reproduces the on-shell matrix
elements
with a current insertion of full QCD, we must add the 
following interaction term 
\be
\ml_j=e^{imvx}\bar{\psi}_c\Gamma \mathcal{Q}B \equiv (\xi^{\dag}_{c,hc}+
\eta^{\dag}_{c,hc}+q^{\dag}_{c,s})j, \qquad j\equiv e^{im_Qvx}
\gamma^0\Gamma\mathcal{Q}B \ ,
\ee
where $B=-G/\sqrt{2}\,\bar l_a\gamma_{\mu}l_b$ in our case,
to the Lagrangian Eq. (\ref{scethcslag}), and perform again the
relevant manipulations (integrating out the small component field
$\eta_{hc}$ and multipole expand), taking now into account the presence 
of the source term. Now the equations of motion lead to
\be
\eta_{c,hc}=\frac{1}{iD_+}\lc gA_{hc+}q_{c,s}-j\rc \ ,
\ee
and inserting this in the effective Lagrangian results in the
following modified source term
\be
\ml_j=\lp \xi_{c,hc}^{\dag}+q_{c,s}^{\dag}+q_{c,s}^{\dag}gA_{hc+}\frac{1}{iD^-}\rp j
\,,
\ee
where the derivative in the last term is hard-collinear and 
acts to the left.
Therefore the full QCD light quark is  matched in the
effective theory into
\be
\psi^{\dag}\to
\xi_{c,hc}^{\dag}+q_{c,s}^{\dag}+q_{c,s}^{\dag}gA_{hc+}\frac{1}{iD^-} \ .
\ee
Putting everything together
results in the following current
\be
J_{QCD}(x)=e^{-im_Qvx}\lc \mathcal{O}_0+ \mathcal{O}_1+\ldots\rc
\,,
\ee
where the terms are labeled with their relevant order
with respect to the dominant term, which reads 
\be
\mathcal{O}_0=\xi^{\dag}_{c,hc}\gamma^0\Gamma Q_v \ .
\ee
Note that we have neglected the term $q_{c,s}^{\dag}\Gamma
Q_v$, since by our kinematical assumptions, there is large momentum
transfer to the final state, so the operators of the effective current
must
contain at least one hard-collinear field to contribute to such final
states. The scaling of each term 
is $\mathcal{O}_0 \sim \lam$ and $\mathcal{O}_1 \sim \lam^2$.
The final conclusion is that up to gauge invariance
subtleties,
the heavy-to-light QCD current is simply matched to
\be
J_{QCD}(x)\to e^{-im_Qvx}\xi^{\dag}_{c,hc}\gamma^0\Gamma Q_v
\equiv  e^{-im_Qvx}\mathcal{O}_0 \ ,
\ee
at tree level. This is the result that we need to study factorization in
heavy-to-light decays at leading order. Hard fluctuations would be included
in the Wilson coefficients $C_k$ of the operators in the effective current
and can be determined by matching calculations so that the general structure 
of the current reads
\be
J_{QCD}(x)=e^{-im_Qvx}\sum_k C_k(m_Q)\mathcal{O}_k \ .
\ee

Using field redefinitions, one can see that in SCETI in
D=4 at leading power the
hard-collinear and soft degrees of freedom decouple
\cite{Bauer:2001yt}. In D=1+1 the same appears to be true. The starting
point is Eq. (\ref{scethcslag}). 
One can see now that redefining the hard-collinear field
$\xi_{c,hc}$ using a soft Wilson line,
\be
\label{redef}
\xi_{c,hc}(x)\equiv Y_s(x)\xi_{c,hc}^{(0)}(x)\ ,
\ee
\be
Y_s(x)\equiv P\exp\lp -ig\int_0^{\infty}ds A_{s-}(x_-+sn_-) \rp \ ,
\ee
and redefining the hard-collinear gluons in the following way:
\be
\label{redef2}
A_{hc}^{\mu}(x)\equiv Y_s(x)A_{hc}^{(0)\mu}Y^{\dagger}_s(x)
\,,
\ee
leads to factorization of hard-collinear and soft modes at
leading order in the effective Lagrangian,
\be
\ml^{(0)}= \bar{\psi}_s
i\Ds_s\psi_s+\xi_{c,hc}^{(0)\dag}
\lp
i\partial^{+}-\frac{m_c^2-i\epsilon}{i\partial^-}
\rp
\xi_{c,hc}^{(0)} 
+\frac{1}{4}
\mathrm{tr}\lc (\partial^+ A^{(0)-}_{hc}-\partial^-A^{(0)+}_{hc})^2\rc
\ .
\ee
The factorization at the level of hard-collinear fields is somewhat 
academic at this stage, since they are not going to appear at leading 
order in the semileptonic decay of the heavy meson. 
Now all the soft-hard collinear dynamics are encoded
in the effective current, that at leading order becomes
\be
\mathcal{O}_0=\xi^{(0)\dag}_{c,hc}Y_s^{\dag}\gamma^0\Gamma Q_v
\,.
\ee

In the following subsection we analyze the semileptonic decay using SCETI. 

\subsection{Semileptonic differential decay rate}

In this section we show how factorization can be implemented in this
process using SCETI. In sec. \ref{SDDR} we have already written the differential decay 
rate in terms of the imaginary part of $T^{--}$.
We can now write this hadronic correlator in terms
of the SCETI fields. The first step consists on the factorization of the hard modes.
We have seen that the QCD current can be expanded in a series
of operators in the effective theory. 
Then the hadronic tensor can be written as
\be
T^{--}\equiv i\int d^2xe^{i(-q+m_Qv)x}\sum_{k=k'+k^{\prime\prime}} H_k(m_Q)
 \la H_Q| T\{\mo_{k'}^{\dag}(x)
\mo_{k^{\prime\prime}} (0)\} | H_Q \ra\equiv \sum_k H_k(m_Q)
T^{--}_{k,eff} \ ,
\ee
where the effective hadronic tensor at leading order is
given by
\be
T^{--}_{0,eff}=i\int d^2xe^{i(-q+m_Qv)x}
 \la H_Q|T\{ \lp \bar \xi_{c,hc} \gamma^- Q_v\rp^{\dag}(x)
 \lp \bar \xi_{c,hc} \gamma^- Q_v\rp(0)\}| H_Q \ra \ .
\ee
After performing the field redefinitions shown in Eq. (\ref{redef}) 
the effective tensor reads
\bea
T^{--}_{0,eff}&=&i\int d^2xe^{i(-q+m_Qv)x}
 \la H_Q| T\{  Q^{\dag}_{v,\alpha}(x) Y(x) Y^{\dagger}(0) Q_{v,\beta} (0)\}| H_Q \ra 
 \\
 \nn
 &&
 \times
 \la 0|T\{(\gamma^-\xi_{c,hc}^{(0)} )_{\alpha}(x) (\bar\xi_{c,hc}^{(0)}\gamma^-)_\beta(0)\}| 0 \ra  \ ,
\eea
where we have used the factorization that the (redefined) soft and hard-collinear modes 
hold at leading order at the Lagrangian level. 
Therefore the correlator can be understood as the convolution of the 
soft and jet function. The heavy quark correlator is explicitely soft gauge invariant 
due to the $Y$ string.  
So far the computation has been pretty much similar to the one in four dimensions 
(in the four-dimensional case there are also hard-collinear strings that here have already 
been approximated to 1). 

Let us now discuss the jet function. We define 
\be
\la 0|T\{(\gamma^- \xi_{c,hc} )_{\alpha}(x) ( \bar\xi_{c,hc}\gamma^-)_{\beta}(0)\}| 0 \ra
\equiv
i\int \frac{d^2k}{2\pi}e^{-ikx}J^-(k)
\,,
\ee
where $\alpha$ and $\beta$ are Dirac indexes. So far we have not specified the quantization frame. 
If we quantize in the equal-time frame we obtain at tree level
\be
\label{jet}
J^-=2\gamma^-\frac{1}{k^+-\frac{m_c^2-i\epsilon}{k^-}}
\equiv
\gamma^-{\tilde J}^{-}
\,.
\ee
If we quantize in the light-front frame with our standard gauge fixing prescription 
$A^+=0$, the $\xi_{c,hc}$ field becomes a constraint (the factor $\frac{m_c^2}{k^-} \sim 
\frac{m_c^2}{m_Q}$ is treated as a correction when quantizing for consistency). 
Either way, the imaginary term 
(the one that appears in the decay rate) reads
(to be kept in mind that $k^- \sim m_Q >0$):
\be
Im J^-=-2\pi\gamma^-\delta(k^+-m_c^2/k^-)
\,.
\ee
The soft function correlator reads
\be
\label{defshape}
\la H_Q|  Q_v^{\dag}\gamma^-(x^-)P[exp(ig\int_0^{x^-}dz^-A^+(z^-))] Q_v(0)| H_Q \ra
\equiv
2\int dp^+ e^{ip^+x^-}S^-(p^+)
\,,
\ee
where $S^-$ is usually named the shape function. 

We are now in position of writing $d\Gamma^{(-)}/dx$ in terms of $S^-$. We 
obtain
\bea
\frac{d\Gamma^{(-)}}{dx}&=&-\frac{1}{M_{H_Q}}\frac{1}{2(2\pi)x}
\frac{G^2}{2\pi}(M_{H_Q}x)^2 
\nn
\\
&&
\times
\int d^2xe^{-i\frac{q^+x^-}{2}}e^{im_Qvx}\int dp^+e^{i\frac{p^+x^-}{2}}
S^-(p^+)\int \frac{d^2k}{(2\pi)2}e^{-ikx}Im {\tilde J}^-
\,.
\label{dGscet}
\eea
By doing the $x^+$ integration and using the fact that the $x^+$ variations of the $Q_v$ field 
are small in comparison with $m_Q$ (already used in the above expression), 
one can set $k^-=m_Q$ in $Im {\tilde J}^-$, which is correct at leading order in the $1/m_Q$ expansion. 

\begin{figure}
\begin{center}
\includegraphics[width=0.49\columnwidth]{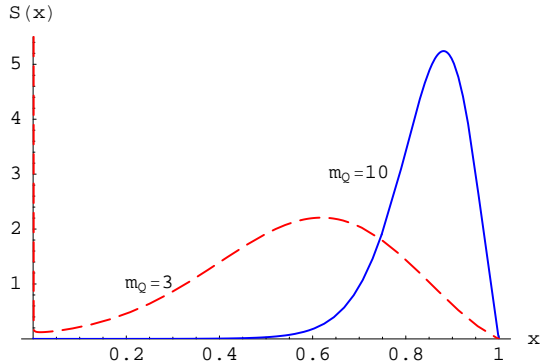}
\caption{\it Plot of the shape function with $x=p^+/P_{H_Q}^+$ for the 
values $m_{s}=\beta$ and $m_Q=10\beta$ (solid line) and $m_{s}=\beta$ and $m_Q=3\beta$ 
(dashed line). Strictly speaking the shape function is singular for $x \rightarrow 0$. 
Nevertheless, this can only be seen by the eye for low values of $m_Q$.}
\label{shapeplot}
\end{center}
\end{figure}

At this stage there is an strong simplification if we work in the light-front 
presented in sec. \ref{QCDLF} with gauge fixing $A^+=0$, since this allows us 
to write the shape function in terms of the $H_Q$-meson wave function squared,
which now reads ($p^++m_Qv^+=xM_{H_Q}$)
\be
\label{shape}
S^-(p^+)
=
\frac{m_Q^2}{M_{H_Q}^2}\frac{\phi_{H_Q}^2(x)}{x^2} \simeq \frac{m_Q^2}{M_{H_Q}^2} \phi_{H_Q}^2(x)
\,.
\ee
Note that we can do that because the shape function is computed at "equal" times 
(i.e. for "$x^+=$ constant") and we can use the free expressions for the 
heavy quark fields. In the last equality we have used the 
fact that terms of order $1-x$ go like $\beta/m_Q$ (at least 
when we compute moments). We  
show how the shape function looks like in Fig. \ref{shapeplot}. It is 
remarkable how similar it is to what one would expect in four dimensions. 

Actually one could do a similar analysis in four dimensions as far as 
one neglects higher Fock components of the $H_Q$ meson state. In this 
approximation one could relate the shape function with the square of 
the $H_Q$ wave function and see what is the impact, for instance, 
in the analysis of Ref. \cite{Beneke:2004in}.  

We could also play the same game for $d\Gamma^{(+)}/dx$. We should then redo 
the construction of the effective theory, since the hard-collinear quark would 
actually go in the opposite direction. 
Everything would 
work analogously changing $J^-$ and $S^-$ by $J^+$ and $S^+$. $J^+$ reads 
equal to $J^-$ changing $\gamma^- \rightarrow \gamma^+$ and $k^+ \leftrightarrow k^-$. 
Therefore $Im{\tilde J}^+$ and $Im{\tilde J}^-$ produce the same delta function 
(as far as the $x$ variable is concerned). The definition of $S^+$ would 
come from Eq. (\ref{defshape}), changing $\gamma^- \rightarrow \gamma^+$ and 
$x^+ \leftrightarrow x^-$. This has consequences when working in the light-front 
frame. If we work in the gauge $A^+=0$ everything is completely analogous to the 
computation of $d\Gamma^{(-)}/dx$, changing $+ \leftrightarrow -$ everywhere. 
In particular this implies that $S^-=S^+$. It is more interesting to consider our 
standard gauge fixing $A^+=0$. In this situation the hard-collinear field is not 
a constraint anymore, but the price to pay is that $S^+$ cannot be easily computed 
because the fields act at different times. Therefore, it is welcome that we can 
obtain $S^+$ from parity arguments.

Our final results for Im$T$ and ${d \Gamma \over dx}$ read
\be
{\rm Im} T^{(tree\,level)}=\pi m_Q^2\phi_{H_Q}^2\lp x +\frac{m_c^2}{m_Q^2}\rp 
\frac{1}{M^2_{H_Q}\lp x+\frac{m_c^2}{m_Q^2}\rp^2}
\,,
\ee
\bea
\label{dGtreelevel}
{d \Gamma \over dx}^{(tree\,level)} &=& 
\frac{1}{M_{H_Q}}\frac{1}{2(2\pi)x} 
\frac{G^2}{2\pi} (M_{H_Q}x)^2 2 {\rm Im} T^{(tree\,level)} 
\\
\nn
&=& 
\frac{G^2M_{H_Q}}{4\pi}
\lp \frac{m_Q}{M_{H_Q}}\rp^2  \frac{1}{\lp x
+\frac{m_c^2}{m_Q^2}\rp^2}\phi_{H_Q}^2\lp x+\frac{m_c^2}{m_Q^2}\rp 
\,,
\eea
where in both expressions, we have included the subleading kinematical 
corrections, and those due to 
the mass of the hard-collinear, which are also parametrically subleading. 

The moments then read
\be
\label{MNscetLO}
M_N^{(tree\,level)}=\frac{G^2M_{H_Q}}{4\pi}\frac{m_{Q}^2}{M_{H_Q}^2}
\lp\frac{m_{Q}^2-m_{c}^2}{m_{Q}^2}\rp^N  
\int_0^1 dx x^{N-2}\phi_{H_Q}^2\lp x\rp
\,.
\ee
We first note that this expression agrees with the result shown in 
Eq. (\ref{MNlayersmeared}) at LO in $1/m_Q$. The $O(m_c^2/m_Q^2)$ 
corrections are also correctly incorporated. The first discrepancies 
are of $O(\beta^2/m_Q^2)$. We need a one-loop analysis to incorporate 
them, which we postpone to the next section.

Finally, we would like to emphasize that the effective field theory computation
has lost the information of the final bound state. 
This can be visualized by comparing Eq. (\ref{dGtreelevel}) with 
the exact hadronic expressions, which consist of a discrete sum over 
resonances.

\section{Soft-(Collinear) Effective Theory in $D=1+1$}

\begin{figure}
\begin{center}
\includegraphics[width=0.49\columnwidth]{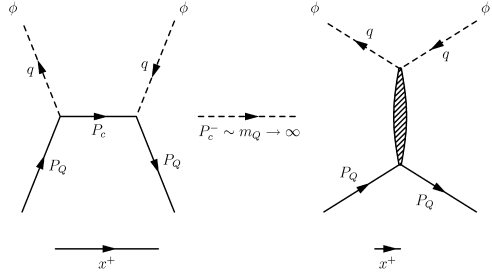}
\caption{\it Symbolic plot to represent the matching from QCD to the effective theory 
of the (tree-level) diagram, the imaginary part of which produces the 
decay of the heavy quark to a hard-collinear quark and the $\phi$ particle. 
The RHS of the figure represents the effective vertex in the effective theory. 
The shape of the effective vertex represents that it is local in $x^+$ but not in $x^-$.}
\label{Bmatching}
\end{center}
\end{figure}

In the previous section we have worked out SCETI and 
the differential decay rate at 
leading order in $\lambda$. The procedure seems to be quite cumbersome. 
In principle, it could be possible 
to extend the previous analysis to higher orders in $\lambda$. Nevertheless, 
the process becomes tedious and, in principle, other modes should be 
included in the theory. This approach does not explicitly profit from the 
fact that hard-collinear modes appear from a very specific interaction 
(the weak one) and that the interaction takes place at very short "times" 
in the $x^+$ axis. Instead we will follow here an alternative approach and we will 
derive the effective theory by integrating 
out any (light) particle with $P^- \sim m_Q$. This will make the effective field 
theory non-hermitian, introducing imaginary terms in the Lagrangian.  
This should not be considered unusual since this also happens for 
non-relativistic effective theories. The idea, which we have already 
mentioned previously, is that diagrams for which the $P^-$ momentum 
flow are of the order of $m_Q$ are local with respect the $x^+$ axis 
in the light-cone frame and, 
once we have chosen the quantization frame to be $x^+=constant$, we 
can construct the corresponding effective Hamiltonian. 
Therefore, the effects due to the hard-collinear fields 
can effectively be reproduced in the effective theory 
by a local vertex in $x^+$ although not local in the other components 
(see Fig. \ref{Bmatching}). We should stress that at the end of the 
day we only want to describe the differential decay rate at leading order 
in the weak interactions. Therefore, they can be described by a purely 
imaginary vertex interaction. 

The effective degrees of freedom of our theory will 
be only soft ones (at least to the order to which we will work here). 
This considerably simplifies our approach compared with others 
where the derivation of the effective theory is performed investigating 
all possible modes existing in the theory. Actually one should not call 
this effective theory "soft-collinear", since 
the collinear modes do not appear in the theory anymore, rather one 
may call it HQET but with some non-hermitian terms.

The fact that our interactions will be almost local in $x^+$-times 
will also have important consequences in how the computation is 
performed. In principle, once all the dependence 
in $P_X^-$ has disappeared from the interaction\footnote{More precisely, 
one can replace $P_X^-$ and $P_X^+$ by a function of $m_Q$, $P_Q^+$ and $q^+$ 
using the equations of motion up to 
residual $x^+$ "time-dependent" terms, which can be expanded in an 
expansion in $1/m_Q$. One could get rid of these time-dependent terms 
systematically using field redefinitions. Nevertheless, in this paper, we 
do not reach enough precision to worry about this problem.}, the interaction only 
depends on $P_Q^+$ and $q^+$. This means that one can use the free-field 
expressions for the fields in the effective interaction. 
Note that this is so even if the momentum $P_X^+$ flowing in the diagram 
is small.
We also 
have to take into account that we know the explicit expressions of the
bound states in terms of the free-field expressions of the field. 
Therefore, we will be able to obtain explicit expressions of the 
matrix elements in terms of the 't Hooft wave function of the bound states,  
once we project 
the effective interacion to the physical states. 
It is then when the non-perturbative dynamics appears. 

An important point here is that the effective vertex can be obtained  
using perturbation theory, order by order in $\beta^2$. Since $\beta^2$ has 
dimensions in two dimensions, the inclusion of $\beta^2$ terms produces 
subleading effects in the OPE. This is so because, even if $P_X^+ \sim \beta$, 
$P_X^+$ typically appears in the combination (except for the leading term) $\beta^2/P_X^2$, 
which could be interpreted as $\sim \beta^2/M_n^2$. 
Therefore, if we restrict ourselves to the kinematical situation $P_X^2 \gg \beta^2$, the 
effective interaction can be obtained using perturbation theory. This is what 
we will do in what follows. 
 
Our starting point is the Lagrangian of QCD where the gluons and 
the $\psi_-$ component have been integrated out, i.e. Eq. (\ref{Lagqcd11}), 
coupled to the electroweak effective vertex (\ref{lagweakhad}). 
We then want to integrate out any (light) degree of freedom 
with $P^- \sim m_Q$ and build an effective theory with only 
soft (light and heavy) quarks. Then we match the 
effective theory onto QCD. 

\begin{figure}
\begin{center}
\includegraphics[width=0.49\columnwidth]{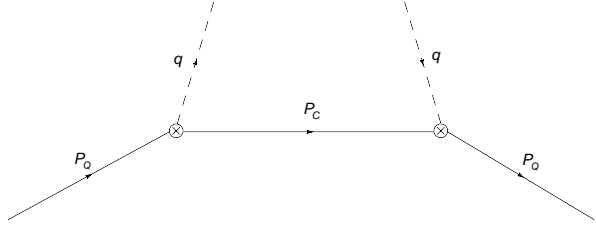}
\caption{\it Tree level diagram, the imaginary part of which 
gives the leading contribution to the effective vertex in the 
effective theory, Eq. (\ref{eqSCET1}).}
\label{SCET1}
\end{center}
\end{figure}

At tree level we only have to compute one diagram, which we have to 
match onto the effective vertex (note that throughout we only demand the 
imaginary piece to be equal). This is simbolically displayed in 
Fig. \ref{SCET1}. 

So far we have not specified neither the gauge nor the quantization frame. For the 
gauge fixing we follow our standard prescription of $A^+=0$. 
Following the notation of the previous section, at lowest order, 
the intermediate hard-collinear field becomes a free particle and its 
propagator reads ($k_{on}^- = m^2/k^+$, $k_{on}^+ = k^+$)
\bea
J^-
&=&
\gamma^-\frac{\xslash{k_{on}}\;\;+m}{k^2-m^2+i\epsilon}\gamma^-
=
\gamma^-\left(
\frac{\gamma^+k_{on}^-}{2}+\frac{\gamma^-k^+}{2}+m
\right)\gamma^-
\frac{1}{k^2-m^2+i\epsilon}
\\
\nn
&=&
2\gamma^-\frac{m^2}{k^2}\frac{1}{k^+-\frac{m^2-i\epsilon}{k^-}}
\equiv
\gamma^-{\tilde J}^{-}
\,,
\eea
if we quantize in the light-front frame, and 
\be
J^-=\gamma^-\frac{\xslash{k} +m}{k^2-m^2+i\epsilon}\gamma^-
=
2\gamma^-\frac{1}{k^+-\frac{m^2-i\epsilon}{k^-}}
\,,
\ee
if we quantize in the equal-time frame. 
We note that the the imaginary term (the one that appears in the decay rate) 
is equal in the light-front or equal-time quantization frame 
(to be kept in mind that $k^- \sim m_Q >0$):
\be
Im J^-=-2\pi\gamma^-\delta(k^+-m_q^2/k^-)
\,.
\ee
Putting everything together, the contribution to the vertex reads (where we 
have used the free equation of motion $k^-=P^-_Q=m_Q^2/P_Q^+$)
\be
{\rm effective\; vertex} \sim {\rm Im} \; 
\left[
\frac{m_c^2}{P_c^2}\frac{m_Q^2}{(P_Q^+)^2}\frac{1}{P_c^+-\frac{m_c^2-i\epsilon}{m_Q^2}P_Q^+}
\right]
\label{eqSCET1}
\,.
\ee

\begin{figure}
\begin{center}
\includegraphics[width=0.49\columnwidth]{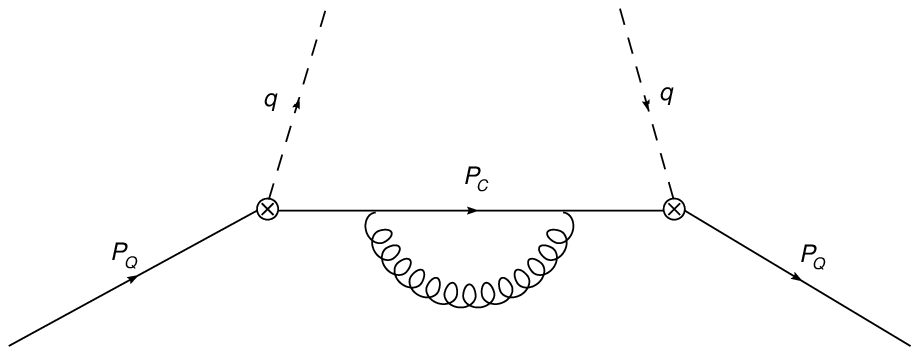}
\includegraphics[width=0.49\columnwidth]{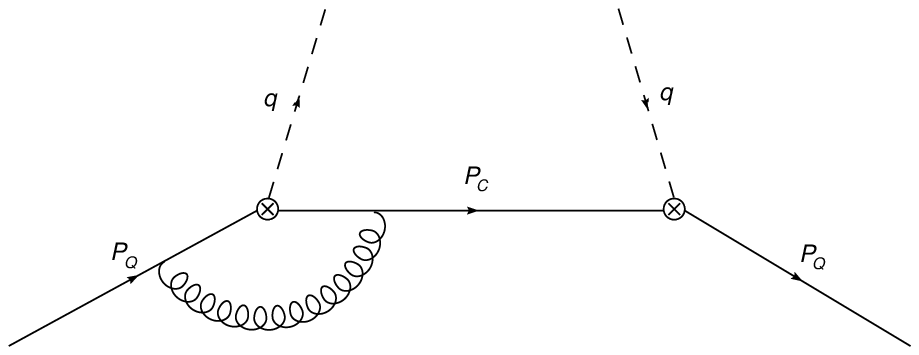}
\caption{\it One loop diagrams (plus their symmetrics), the imaginary part of which 
contribute to the effective vertex in Eq. (\ref{eqSCET2}).}
\label{SCET2}
\end{center}
\end{figure}

At $O(\beta^2)$, the following diagrams have to be 
considered (Fig. \ref{SCET2}). Their net effect is to renormalize the masses 
of the hard collinear and heavy quark. The first diagram renormalizes the hard-collinear 
mass that appears in the hard-collinear propagator. The effect of the second 
diagram is to renormalize the masses of the vertex. To simplify the expression
we keep subleading terms in the $1/m_Q$ expansion. 

By adding all the terms, the effective interaction can be written as ($P_c^+=P_Q^+-q^+$)
\bea
\label{eqSCET2}
{\rm effective\; vertex} &\sim& {\rm Im} \; 
\left[
\frac{m_{c,R}^2}{P_c^2}\frac{m_{Q,R}^2}{(P_Q^+)^2}\frac{1}{P_c^+-\frac{m_{c,R}^2-i\epsilon}{m_{Q,R}^2}P_Q^+}
\right]
\\
\nn
&=&
-\pi \frac{m_{Q,R}^2}{(P_Q^+)^2}\delta \left(\left(1-\frac{m_{c,R}^2}{m_{Q,R}^2}\right)P_Q^+-q^+\right)
\,,
\eea
where with the precision of the calculation, we have replaced $m_Q^2$ by $m_{Q,R}^2$ within the 
delta.

We are now in the position to write the effective Lagrangian, which, in fact, is the 
HQET Lagrangian adding the effective vertex. 
 
\be
{\cal L}={\cal L}_{HQET}+Im[{\cal L}_I]
\ee
and the effective vertex reads
\be
\label{LI}
\mathcal{L}_I=
-\frac{G^2}{2\pi}(\partial^+\phi)\lp \frac{m_{Q,R}}{i\partial^+}Q_{+}\rp^{\dagger}\frac{1}{i\partial^+-\frac{m_{c,R}^2-i\epsilon}{m^2_{Q,R}}
i\partial^+}\lp 
\frac{m_{Q,R}}{i\partial^+}Q_{+}\rp(\partial^+\phi^{\dagger}) 
\,.
\ee
In order to keep the expression for the effective vertex more compact, we have written it in terms 
of the field $Q_+$, had we written it in terms of $Q_{+_v}$, there would be a shift in the derivatives 
(for instance $i\partial^+ \rightarrow m_Qv^++i\partial^+$). We note that the 
Lagrangian is local in $x^+$. 

\subsection{Semileptonic differential decay rate}
\label{comparison}

At this stage we can compute the semileptonic decay. 
Actually we will have to compute the imaginary part produced by the 
effective vertex (\ref{LI}):
\be
{\rm Im} T_{eff}=\pi m_{Q,R}^2\phi_{H_Q}^2\lp \frac{x}{1
-\frac{m_{c,R}^2}{m_{Q,R}^2}}\rp \frac{1}{(M_{H_Q}x)^2}\lp 1-\frac{m_{c,R}^2}{m_{Q,R}^2}\rp
\,.
\ee
The differential decay rate then reads
\bea
\label{dG}
{d \Gamma \over dx}^{pert} &=& \frac{1}{M_{H_Q}}\frac{1}{2(2\pi)x} \frac{G^2}{2\pi} 
(M_{H_Q}x)^2 2 {\rm Im} T_{eff} 
\\
\nn
&=& 
\frac{G^2M_{H_Q}}{4\pi}\frac{m_{Q,R}^2-m_{c,R}^2}{m^2_{Q,R}}\lp 
\frac{m^2_{Q,R}}{M^2_{H_Q}}\rp  \frac{1}{x}\phi_{H_Q}^2\lp \frac{x}{1-\frac{m_{c,R}^2}{m_{Q,R}^2}}\rp 
\,.
\eea
Let us note that we can also rewrite this result (actually one equivalent with the precision of our calculation) 
in terms of shape and jet functions as in Eq. (\ref{dGscet}). 
For the shape function, Eq. (\ref{shape}), we have to replace $m_Q \rightarrow m_{Q,R}$ and for the jet 
function, Eq. (\ref{jet}), we have to replace $m_c \rightarrow m_{c,R}$.

The expressions for the moments read
\be
\label{MNscet}
M_N^{pert}=\frac{G^2M_{H_Q}}{4\pi}\frac{m_{Q,R}^2}{M_{H_Q}^2}\lp\frac{m_{Q,R}^2-m_{c,R}^2}{m_{Q,R}^2}\rp^N  
\int_0^1 dx x^{N-2}\phi_{H_Q}^2\lp x\rp
\,.
\ee
This expression allows us to compare the hadronic and the effective field theory computations. 
It is quite remarkable that this result agrees with the hadronic result obtained in Eq. (\ref{MNlayersmeared}). 
Therefore, we also obtain the same expressions for $M_N^{OPE}$, Eq. (\ref{MNlayerexp}), valid in the OPE kinematic 
region, and for $M_N^{SCETI}$, Eq. (\ref{MNlayerscetexp}), valid in the SCETI kinematic region.
The comments about the precision of the 
expressions  made there also apply 
here. The expression for $M_N^{OPE}$
 is correct up to, and including, $O(1/m_Q^2)$ corrections.
The expression for $M_N^{SCETI}$ is correct up to, and including, $O(1/m_Q)$ corrections. 
The conclusion is that, if working with moments, we do not see duality violations 
with the precision of our calculation. 

It is interesting to compare with the computation of the moments made by 
Bigi et al. \cite{Bigi:1998kc}.
For the total decay width, $M_1$, we obtain exactly the same analytic expression. 
Actually, this result is not affected by the radiative corrections 
(we obtain the same result at tree level or one-loop). 
In our opinion, this explains the agreement, since the computation of Bigi et al. has  
(effectively) been done at tree level. For $M_{0,2}$, we need to perform the one-loop computation in order 
to get agreement with their results with $O(1/m_Q^2)$ precision. This is quite remarkable, since 
in the kinematical situation they choose, $q^+=0$, it is argued that there is a 
non-renormalization theorem for the current, whereas in the 
kinematical situation $q^-=0$ this is not true and the one-loop 
corrections to the vertex shown in Fig. \ref{SCET2} have to be included.

We are now in the position to perform a numerical analysis of the corresponding 
expressions obtained from effective field theories with perturbative factorization 
and to compare them with the hadronic ones. We first consider the differential 
decay rate. As we have already discussed throughout the paper, the direct 
comparison between the partonic and hadronic result is not possible since 
the first is an smooth function in $x$, whereas the second consists of a 
sum over deltas. Nevertheless, the layer-function approximation gives us a
qualitative way to compare the hadronic matrix elements with the computation 
using the effective theory, since it allows us to write the hadronic matrix elements 
in terms of only the $H_Q$ meson wave-function. Naively, one could make the following 
assignment to the hadronic matrix element from the effective field theory computation 
\bea
\label{difratescet}
&&
\pi\beta\frac{m_{Q,R}}{M_{H_Q}^2}\left(1-\frac{m_{c,R}^2}{m_{Q,R}^2}\right)
\frac{1}{x}\phi_{H_Q}\lp \frac{x}{1-\frac{m_{c,R}^2}{m_{Q,R}^2}}\rp 
\\
\nn
&&
\quad
\simeq
\pi\beta\frac{m_{Q,R}}{M_{H_Q}^2}
\frac{1}{x}\phi_{H_Q}\lp x\rp 
\left[1+2x\frac{m_{c,R}^2}{M_{H_Q}^2}
\left(\frac{\phi_{H_Q}^{\prime}(x)}{\phi_{H_Q}(x)}-\frac{1}{2x}\right)
\right]
\,,
\eea
where the normalization has been adjusted to agree with the layer-function result. 
The LO result agrees with the LO layer-function expression, Eq. (\ref{optionBR}). Nevertheless, this is not so
for the subleading one. We will elaborate on this issue at the end of this section.
Although numerically the effect is not very important (we would obtain very similar plots to 
those obtained in Figs. \ref{difrate} and \ref{difratemQ}), it is important 
from the conceptual point of view.

\begin{figure}
\begin{center}
\includegraphics[width=0.49\columnwidth]{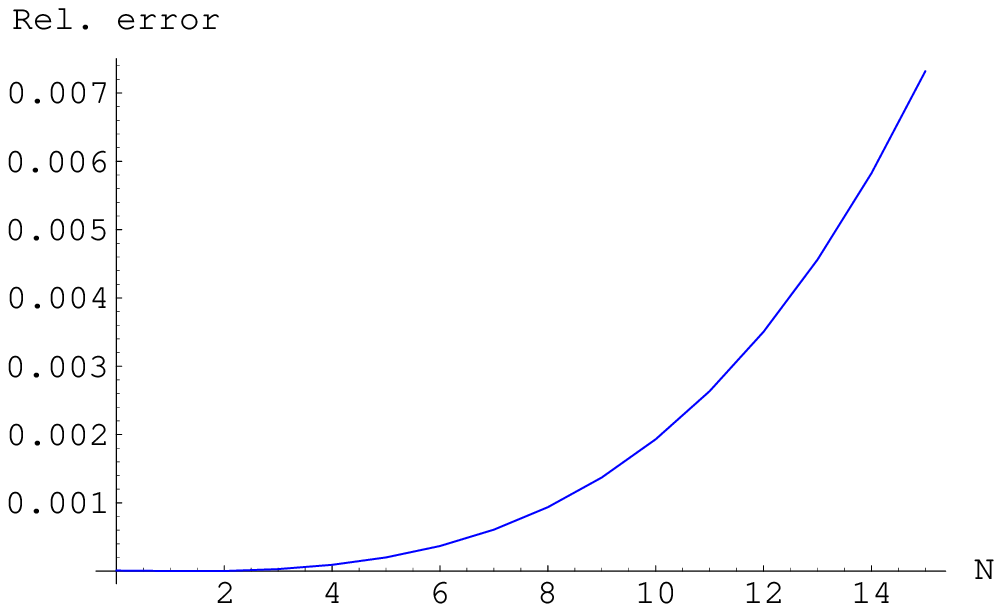}
\includegraphics[width=0.49\columnwidth]{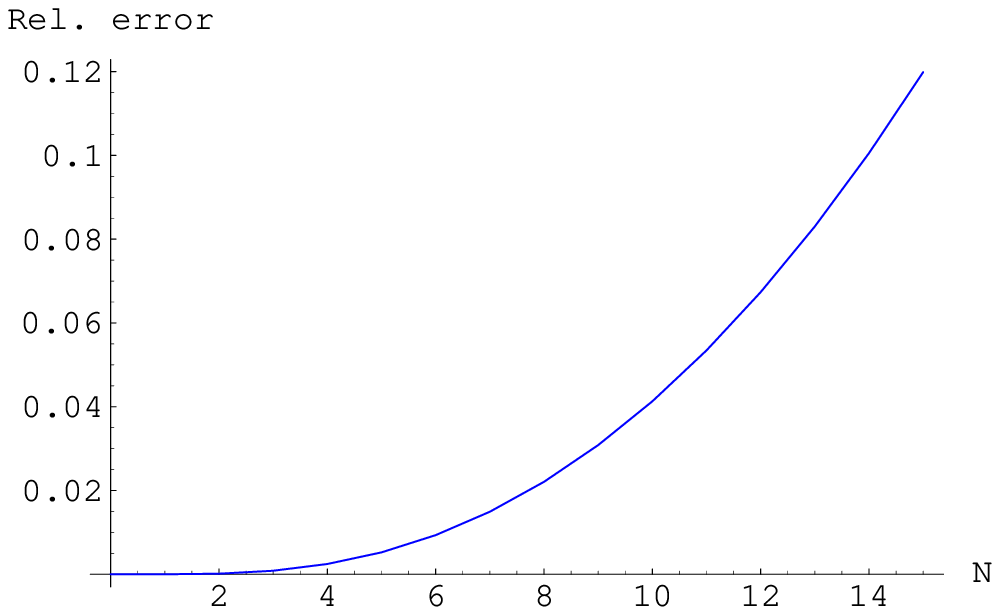}
\includegraphics[width=0.49\columnwidth]{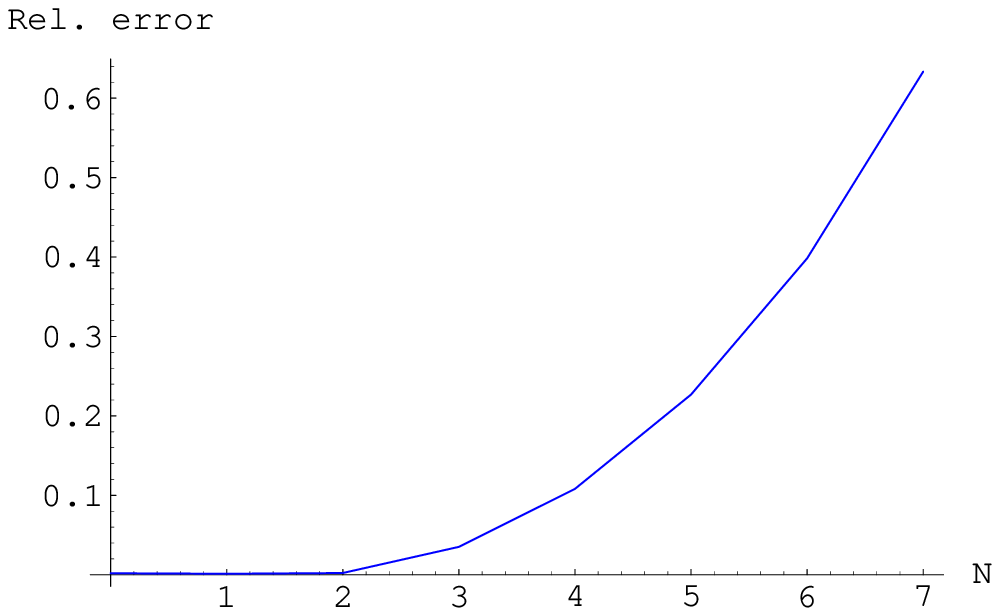}
\caption{\it Difference between the hadronic, Eq. (\ref{MNhadrBigi}), 
and perturbative, Eq. (\ref{MNscet}), result for the 
moments (divided by the hadronic result). The first figure is for the values $m_Q=10\beta$, $m_c=\beta$ and $m_s=\beta$, 
the second is for the values $m_Q=15\beta$, $m_c=10\beta$ and $m_s=0.56\beta$ and the third for the 
values $m_Q=3\beta$, $m_c=\beta$ and $m_s=\beta$.} 
\label{figmoments}
\end{center}
\end{figure}

\begin{figure}
\begin{center}
\includegraphics[width=0.49\columnwidth]{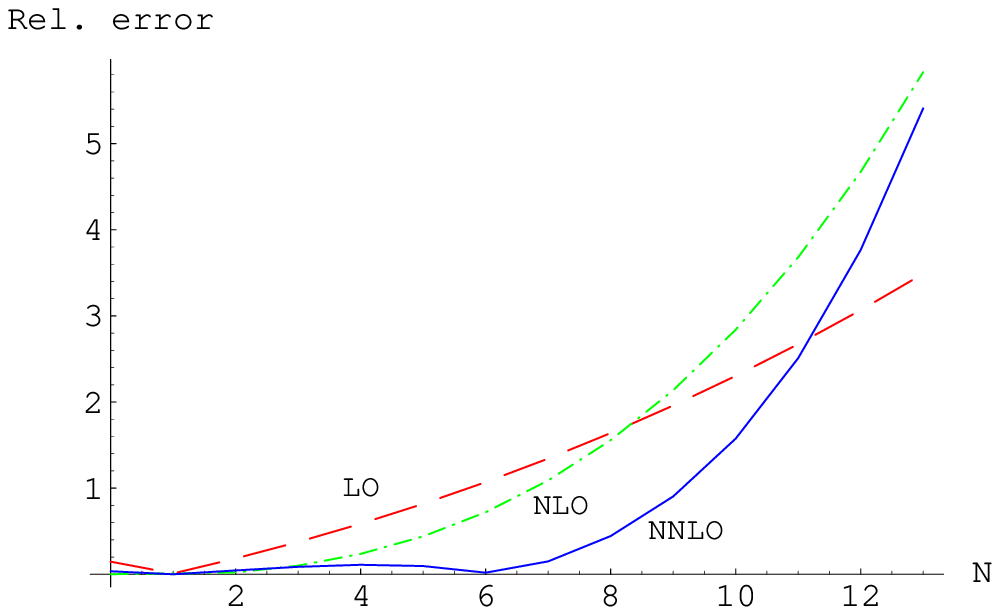}
\includegraphics[width=0.49\columnwidth]{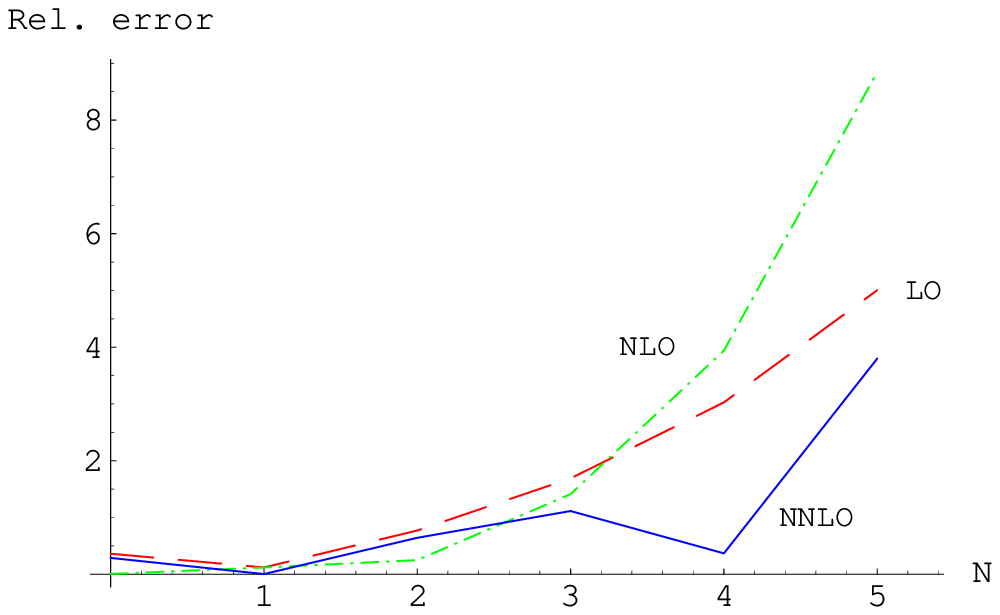}
\caption{\it Difference between the hadronic, Eq. (\ref{MNhadrBigi}), 
and perturbative, Eq. (\ref{MNlayerexp}), result for the 
moments (divided by the hadronic result) in the OPE limit. The dashed line 
for the LO result, the dash-dotted line for the NLO result, and the solid line 
for the NNLO result. We take the 
values $m_Q=10\beta$, $m_c=\beta$ and $m_s=\beta$ for the first figure and 
$m_Q=3\beta$, $m_c=\beta$ and $m_s=\beta$ for the second figure. We use the values  
$\langle t\rangle=1.73\beta$ and $\langle t^2\rangle=3.99\beta^2$, which can 
be checked with the sum rules of Ref. \cite{Burkardt:2000ez}.} 
\label{momOPE}
\end{center}
\end{figure}

\begin{figure}
\begin{center}
\includegraphics[width=0.49\columnwidth]{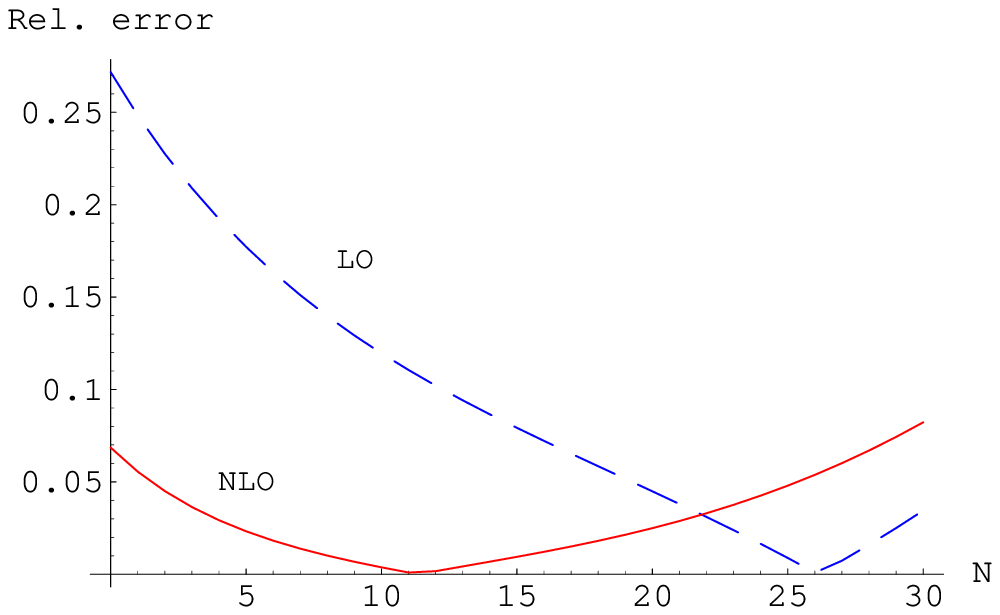}
\includegraphics[width=0.49\columnwidth]{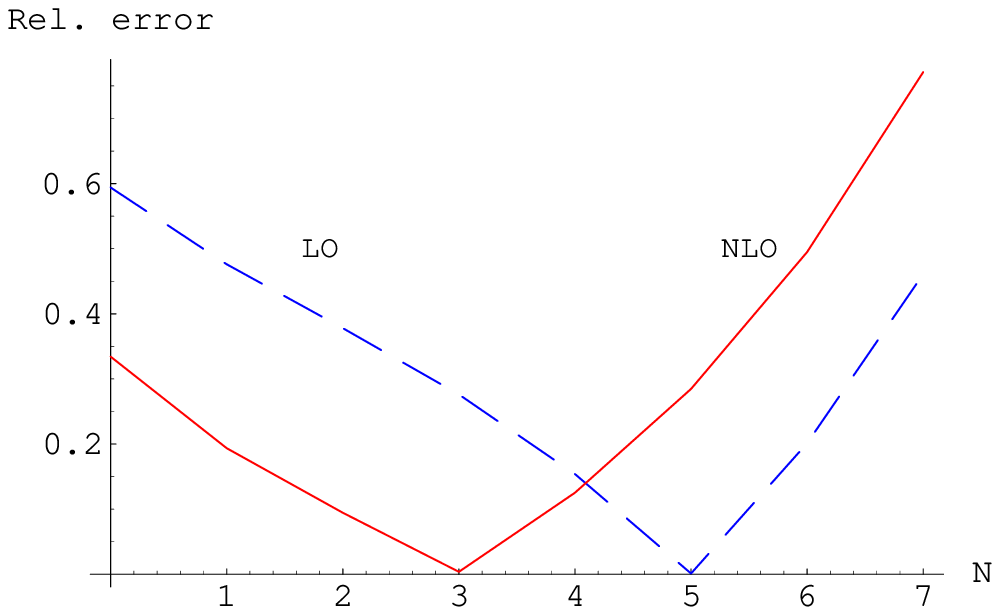}
\caption{\it Difference between the hadronic, Eq. (\ref{MNhadrBigi}), 
and perturbative, Eq. (\ref{MNlayerscetexp}), result for the 
moments (divided by the hadronic result) in the SCETI limit. The dashed line 
for the LO result,  and the solid line 
for the NLO result. We take the 
values $m_Q=10\beta$, $m_c=\beta$ and $m_s=\beta$ for the first figure and 
$m_Q=3\beta$, $m_c=\beta$ and $m_s=\beta$ for the second figure. } 
\label{momSCET}
\end{center}
\end{figure}

We now move to the comparison of the moments. In this case, the comparison is more 
sound. For all the numerical checks we have set $G^2/(4\pi)=1$. 
We have compared the perturbative and hadronic result. The agreement is 
very good up to very high moments. We show this comparison in Fig. \ref{figmoments}\footnote{
We also make the comparison for the set of masses used in Ref. \cite{lebed}, to illustrate 
how well our expressions would work for an hypothetical $b \rightarrow c l\nu$ decay.}. 
We also consider how convergent is the expansion either if we work in the strict 
OPE or SCETI regime. We show the convergence of the 
OPE expansion in Fig. \ref{momOPE}. For the OPE limit we can see that the 
breakdown of the agreement with the hadronic result appears earlier. This is 
to be expected since there is a new scale $N/m_Q$, which is not resummed. In
any case the precision is better than 5\% for $N$ below 6 at NNLO. We also 
perform the analysis in the SCETI region. We show the plot in Fig. \ref{momSCET}. 
As expected the convergence improves over the OPE evaluations and they are 
optimal for values of $N\sim m_Q/\beta$ as expected. For very large values of $N$, 
they blow up as expected. We also consider the dependence on the heavy quark mass 
of these results. For that we repeat the same analysis for $m_Q=3$, which 
is somewhat the limiting case of validity of our results. Overall, we 
get a similar picture than before but with worse convergence (actually in the OPE 
region the results are barely convergent) and the perturbative 
results are only reliable for lower moments, again as expected. The effect of the 
one-loop corrections (the renormalization of the masses) is small.

Keeping 
the prefactor $\displaystyle{\lp\frac{m_{Q,R}^2-m_{c,R}^2}{m_{Q,R}^2}\rp^N}$
improves the numerical agreement with the hadronic expression but we should 
remind that we cannot claim better accuracy than $\simeq 1-N \frac{m_{c,R}^2}{m_{Q,R}^2}$.
This effect is particularly important if the mass of the hard-collinear is large.

\begin{figure}
\begin{center}
\includegraphics[width=0.49\columnwidth]{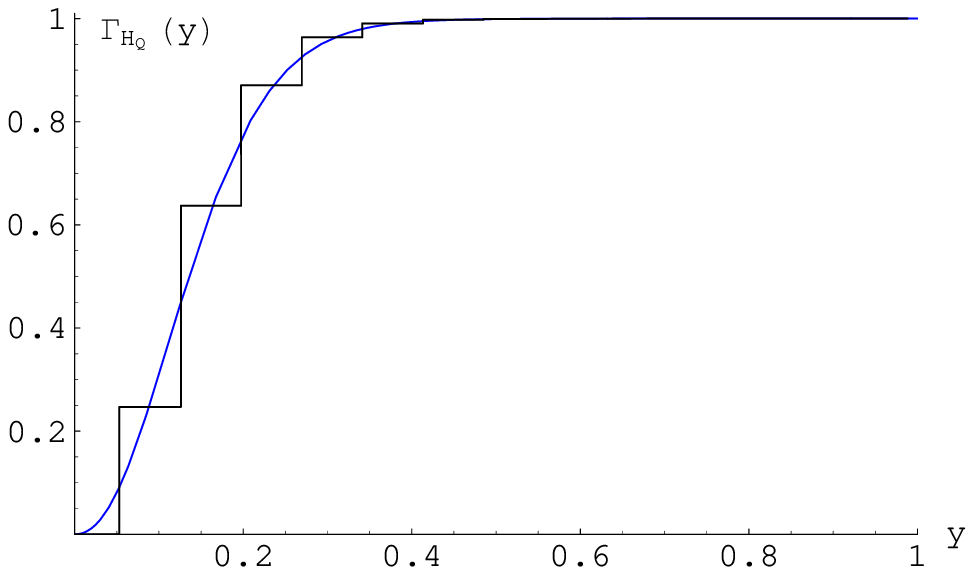}
\includegraphics[width=0.49\columnwidth]{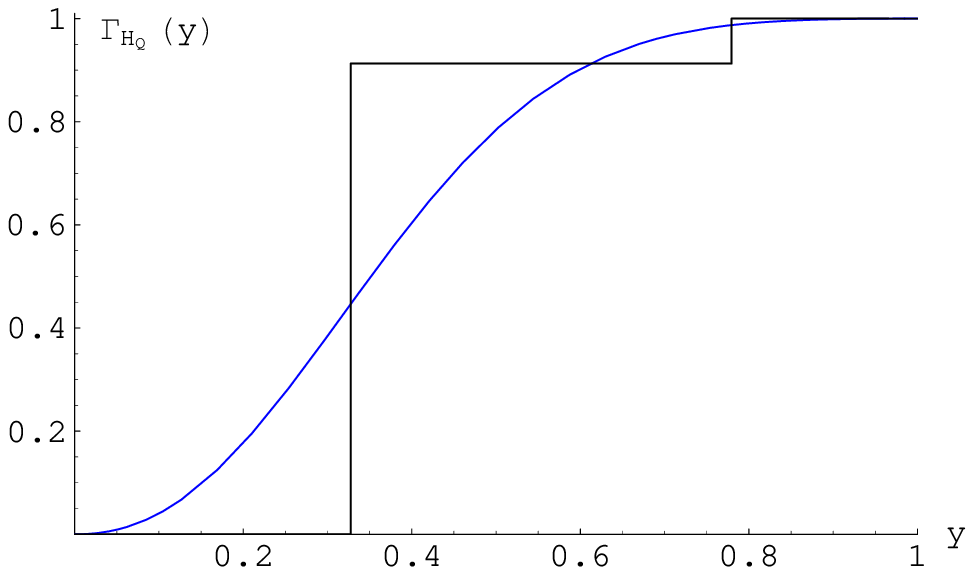}
\caption{\it Plot of $\Gamma_{H_Q}(y)$ using the hadronic or 
perturbative expression of $d\Gamma/dx$. The  smooth curve represents the 
perturbative prediction and the other line is
 the hadronic one.  We take the 
values $m_Q=10\beta$, $m_c=\beta$ and $m_s=\beta$ for the first figure and 
$m_Q=3\beta$, $m_c=\beta$ and $m_s=\beta$ for the second figure. } 
\label{figdelta}
\end{center}
\end{figure}

We now consider $\Gamma_{H_Q}(y)$. We compare the hadronic result
versus the prediction from effective field theories in Fig. \ref{figdelta}. 
We can see a good quantitative agreement between both lines but it is not 
possible to perform a point-to-point comparison because the hadronic result 
is a step-function whereas the perturbative computation is an smooth function 
in $y$. Therefore, it is difficult to quantify the error, which typically is 
of the order of the difference between the $n$ and $n+1$ state contribution to 
$\Gamma_{H_Q}(y)$. 

To make more evident this problem, we now consider the average
\be
\int_{x_n-\delta x}^{x_n+\delta x}\frac{d \Gamma}{dx} dx
\,,
\ee
where $\delta x$ is bounded to be small enough that only one resonance 
contributes to the integral. One could believe that the duality violations can be smoothed in this 
way. This definition provides the closest possible thing to a point-to-point 
comparison between the perturbative and hadronic result. Since we are now able to perform analytic computations both for the 
hadronic and for perturbative result, we are in the position to quantify 
this statement (we restrict to the kinematic regime where $M_n^2 \gg \beta^2$). 
We obtain the following result from the hadronic computation
\be
\int_{x_n-\delta x}^{x_n+\delta x} \frac{d \Gamma}{dx} \Bigg|_{hadr.}dx
=\Gamma_n \simeq
\frac{\pi^2 \beta^2}{M^2_{H_Q}}
\frac{G^2M_{H_Q}}{4\pi}
\frac{m_{Q,R}^2}{M^2_{H_Q}}
\frac{1}{x_n}
\phi_{H_Q}^2\left(x_n\right)
\left[1+2\frac{m_{c,R}^2}{M_{H_Q}^2}
\left(\frac{\phi_{H_Q}^{\prime}(x_n)}{\phi_{H_Q}(x_n)}-\frac{1}{x_n}\right)
\right]
\,.
\ee
From the perturbative computation we obtain
\be
\int_{x_n-\delta x}^{x_n+\delta x} \frac{d \Gamma}{dx} \Bigg|_{pert.}dx
\simeq 
2\delta x
\frac{G^2M_{H_Q}}{4\pi}
\frac{m_{Q,R}^2}{M^2_{H_Q}}
\frac{1}{x_n}
\phi_{H_Q}^2\left(x_n\right)
\left[1+2x_n\frac{m_{c,R}^2}{M_{H_Q}^2}
\left(\frac{\phi_{H_Q}^{\prime}(x_n)}{\phi_{H_Q}(x_n)}-\frac{1}{2x_n}\right)
\right]
\,.
\ee
We can see that both expressions are different, even at leading order, due to the 
normalization. Only 
if we fine tune $\delta x$ to a very specific value (actually, the one chosen in Ref. \cite{lebed}), 
we get agreement between both expressions, 
and, even then, the subleading corrections are different\footnote{Actually, these 
small differences for the subleading corrections are crucial to get agreement for the moments at $O(1/m_Q^2)$.}. 
In order to fine 
tune the value of $\delta x$, we should have a good knowledge of the non-perturbative 
spectrum (this is certainly difficult in the four-dimension case, although one 
can always assume a linear regge behavior), which is something that we do not expect 
can be achieved from perturbation theory. On the other hand, as far as $\delta x$ is 
independent of $n$, one obtains this equality at leading order (subleading corrections 
remain to be different)
\be
\displaystyle{\frac{\int_{x_{n+1}-\delta x}^{x_{n+1}+\delta x} \frac{d \Gamma}{dx} \big|_{hadr.}dx}
{\int_{x_n-\delta x}^{x_n+\delta x} \frac{d \Gamma}{dx} \big|_{hadr.}dx}}
=
\frac{\int_{x_{n+1}-\delta x}^{x_{n+1}+\delta x} \frac{d \Gamma}{dx} \big|_{pert.}dx}
{\int_{x_n-\delta x}^{x_n+\delta x} \frac{d \Gamma}{dx} \big|_{pert.}dx}
\ee
without fine-tuning $\delta x$.

\section{Conclusions}

We have studied QCD in 1+1 dimensions in the large $N_c$ limit 
using light-front Hamiltonian perturbation theory in the 
$1/N_c$ expansion. We have used the formalism developed to 
exactly compute hadronic transition matrix elements for arbitrary currents 
at leading order in $1/N_c$. We have  
compared with previous results found in the literature. 
We have computed in two alternative ways the semileptonic differential 
decay rate of a heavy meson and its moments using the previously computed hadronic matrix 
elements. They yield very different expressions for the 
differential decay rate, which should be equal by parity invariance. This has 
lead to the derivation of non-trivial equalities between matrix elements, which 
we have checked numerically. Some partial analytic checks have also been done.

We have then focused on the kinematic regime where the OPE ($N \sim x \sim 1$) 
or SCETI ($1-x \sim 1/N \sim \beta/m_Q$) can be applied. 
This means $ n \gg 1$, where $n$ is the principal quantum number of the final hadronic state. 
This has allowed us to use the properties of the final hadronic bound state in those kinematical 
regimes by using the layer function \cite{einhorn,Brower:1978wm}. The resulting expressions are suitable 
to a more direct connection with the computation obtained using perturbative factorization. We 
have then obtained expressions for the moments using the EulerMcLaurin formula, within 
an expansion in $1/m_Q$, either in the OPE  or in 
the SCETI regions.
In the first case up to order $1/m_Q^2$ and in the second up to order $1/m_Q$. 
We have also checked that these results agree with the expressions of the sum rules for $N=0,1,2$ obtained 
in Ref. \cite{Bigi:1998kc} up to order $1/m_Q^2$ (we have also checked them by direct 
computation).

We have also studied the differential decay rate using effective theories with 
perturbative factorization. We have first derived SCETI at leading order in 1+1 dimensions, 
and applied it to the differential decay rate. We have seen that there is an strong 
simplification working in the light-cone quantization frame, which allows us to relate the shape function 
with the wave-function of the bound state (in the large $N_c$). We have then dwelt further 
on the issue of finding the optimal effective theory to describe the differential decay rate
in the OPE and SCETI kinematical regime. We see that, at least in two dimensions, it appears to be
more efficient to integrate out the hard-collinear modes and only keep soft degrees of 
freedom. This takes advantage of the fact that the hard-collinear interactions only appear in 
a very specific way, emanating from the weak interaction, whereas in the standard 
construction of SCETI one works out all possible couplings. The resulting effective theory 
is equal to HQET plus some imaginary terms, which describe the differential decay rate. 
This effective theory is more efficiently implemented in the light-front quantization frame, 
where it becomes "local" in the "$x^+$" quantization frame, and it is suitable for computations 
in a Hamiltonian formulation. We have then obtained the differential decay rate at one-loop.
We would also like to remark that some of these ideas could also apply to the four-dimensional 
case, where it also seems possible to obtain a "local" effective interaction in the 
"$x^+$" quantization frame.

At the end of the day, we have been able to obtain expressions for the moments using 
effective field theories with perturbative factorization with relative accuracy of 
$O(\beta^2/m_Q^2)$ in the kinematical regime where the OPE can be applied, and 
with relative accuracy of $O(\beta/m_Q)$ in the kinematical regime where non-local 
effective field theories can be applied. These expressions agree, within this precision, 
with those obtained from the hadronic expressions using the layer-function 
approximation plus Euler-McLaurin expansion (taking into account the numerical 
agreement for the radiative correction).  

Numerically very good agreement for the moments between the exact result and 
the result using effective field theories with perturbative factorization is obtained.
For the differential decay rate it is also possible to perform a sort of comparison 
between the hadronic and perturbative result based on the layer-function approximation. The 
agreement is very good if we do not approach too much the limit $x \rightarrow 1$. 

There is still the issue of the theoretical error. As we have mentioned throughout the 
paper, it is not possible to make a quantitative comparison (with errors) 
between the hadronic and perturbative differential decay rate, since one is represented 
by a sum of deltas whereas the other is a smooth function in $x$. Therefore, 
effective field theories with perturbative factorization are not suitable to 
predict the differential decay rate on a point-to-point basis. 
Note that 
this comment also applies to $\Gamma_{H_Q}(y)$ and, in principle, 
to other arbitrarily smeared functions. This is best 
illustrated in the derivation of SCETI at leading order where, after field 
redefinitions, one is lead to an effective theory where the hard-collinear 
is a free field. Therefore, it can never build a bound state, which is 
what is observed in the hadronic differential decay rate. Effective 
field theories with perturbative factorization can only hopefully be a 
good aproximation (or at least a good starting point) for inclusive 
observables on which one averages over a large fraction of the final bound 
states (this is a handicap from the experimental point of view, since 
one has to know the differential decay rate for arbitrarily large momenta). 
This motivates the use of moments\footnote{When we perform the comparison 
among moments the difference is "less severe" than 
using directly the differential decay rate. The point is to quantify the "less severe" 
at the parametrical level (although it can be done at the numerical level). 
For the moments $N=0$, 1, 2, this has already been done in Ref. \cite{Bigi:1998kc}.} 
for the comparison, since they may lead to 
a "quantitative" comparison between hadronic and perturbative results. With the precision 
obtained in this paper we obtain a perfect match between the hadronic and 
perturbative results. Nevertheless, to have a more rigourous handle of the 
errors, one should be able to quantify the error produced by using the layer-function 
(i.e. to consider subleading effects in the WKB approximation), as well as by using 
the Euler-McLaurin formula. Actually it is this last formula that allows to 
make quantitative the comparison of perturbative and non-perturbative results. 
We expect to come back to these issues in the future in order to try to find 
duality violations in the computation of the moments in the SCETI kinematical 
region.


\vspace{5mm}
\noindent
{\bf Acknowledgments:}\\
The work of J.M. and A.P. was supported in part by a {\it Distinci\'o} 
from the {\it Generalitat de Catalunya}, as well as by the contracts 
MEC FPA2004-04582-C02-01, CIRIT 2005SGR-00564 and 
the EU network EURIDICE, HPRN-CT2002-00311. The work of J.R. was supported in 
part by the contract MEC FPA2002-2415.

\appendix

\section{Conventions and notation}
\label{notation}
In this appendix the conventions and notation that we use are
presented.
We define two light-like vectors 
(with the metric $g^{+-}=g^{-+}=2$ and zero elsewhere), 
\be
n_-^{\mu}=(1,1), \quad   n_+^{\mu}=(1,-1)
\,,
\ee
where light-cone coordinates are defined in the usual way,
\be
x^+\equiv n_+\cdot x=\lp x^0+x^1\rp, 
\quad
x^-\equiv n_-\cdot x=\lp x^0-x^1\rp, \quad
\ee
which imply that
\be
x^0\equiv\frac{1}{{2}}\lp x^++x^-\rp, \quad
x^1\equiv\frac{1}{{2}}\lp x^+-x^-\rp, \quad
\ee
and
\be
\partial^-=2{\partial \over \partial x^+} 
=
{\partial \over \partial x^0}+{\partial \over \partial x^3}
=\partial_0+\partial_3 \sim p^- \,,
\partial^+=2{\partial \over \partial x^-} 
=
{\partial \over \partial x^0}-{\partial \over \partial x^3}
=\partial_0-\partial_3\sim p^+
\,,
\ee
\be
P\cdot x= \frac{P^+x^-}{2}+\frac{P^-x^+}{2}
\,,
\ee
\be
d^Dx={1\over 2}dx^+dx^-d^{D-2}x_{\bot}
\,.
\ee
For the Dirac algebra is useful to define the corresponding light-cone 
matrices
\be
\ns_+=\gamma^+, \quad \ns_-=\gamma^-
\,.
\ee
To have explicit expressions, it is useful to work with
an explicit representation of the Dirac algebra.
We will use the following Weyl-like representation for the Dirac algebra
\be
\gamma^0=\lp \begin{array}{cc} 0 & -i \\ i & 0\end{array}\rp \quad
\gamma^1=\lp \begin{array}{cc} 0 & i \\ i & 0\end{array}\rp
\,,
\ee
so that the corresponding light cone matrices are given by
\be
\gamma^-=-2i\lp \begin{array}{cc} 0 & 1 \\ 0 & 0\end{array}\rp \quad
\gamma^+=2i\lp \begin{array}{cc} 0 & 0 \\ 1 & 0\end{array}\rp
\,.
\ee
This way one can see the explicit effect of the projection
operators ($\gamma_5=\gamma^0\gamma^1$)
\bea
\Lambda_+&\equiv& \frac{1+\gamma_5}{2}=\frac{\gamma^0\gamma^+}{2}=
\frac{1}{4}\ns_-\ns_+=\frac{1}{4}\gamma^-\gamma^+
= \lp \begin{array}{cc} 1 & 0 \\ 0 &
  0 \end{array}\rp
\,,
\\
\nn
\Lambda_-&\equiv& \frac{1-\gamma_5}{2}=\frac{\gamma^0\gamma^-}{2}
=\frac{1}{4}\ns_+\ns_-
=\frac{1}{4}\gamma^+\gamma^-=\lp  \begin{array}{cc} 0 & 0 \\ 0 &
  1\end{array}\rp
\,,
\eea
in the sense that if one splits a quark doublet on its two components
\be
\psi= \lp \begin{array}{c} \psi_+ \\ \psi_- \end{array}\rp
\,,
\ee
then the projection operators act as expected
\be
\Lambda_+\psi=\lp \begin{array}{c} \psi_{+} \\ 0 \end{array}\rp \quad
\Lambda_-\psi=\lp \begin{array}{c} 0 \\ \psi_- \end{array}\rp
\,.
\ee


\end{document}